\documentclass[reprint,amsmath,amssymb,aps]{revtex4-2}

\usepackage{graphicx}
\usepackage{amsmath,amssymb,amsthm} 
\usepackage{leftidx}
\usepackage{hyperref}
\hypersetup{colorlinks,allcolors=black}
\usepackage{floatrow}
\DeclareUnicodeCharacter{2212}{-}

\newcommand{\old}[1]{}

\begin{document}

\preprint{APS/123-QED}

\title{Oscillation properties of relativistic tori in the vicinity of a distorted deformed compact object}
\author{Shokoufe Faraji}
 \email{shokoufe.faraji@zarm.uni-bremen.de}
\author{Audrey Trova}%
 \email{audrey.trova@zarm.uni-bremen.de}
\affiliation{%
 University of Bremen, Center of Applied Space Technology and Microgravity (ZARM), 28359 Germany
}%
%


\begin{abstract}

This paper studies the oscillation properties of relativistic, non-self-gravitating tori in the background of a distorted deformed compact object. This work concentrates on the static and axially symmetric metric containing two quadrupole parameters; relating to the central object and the external fields. This metric may associate the observable effects to these parameters as dynamical degrees of freedom. The astrophysical motivation for choosing such a field is the possibility of constituting a reasonable model for an actual scenario occurring in the vicinity of compact objects. This paper aims to investigate the radial epicyclic frequency in a perfect fluid disk and not a test particle scenario via a local analysis. To achieve this goal, we employ the vertically integrated technique to able to treat the equation analytically. The tori are also modelled with Keplerian and non-Keplerian distributions of specific angular momentum, and we discuss the dependence of oscillation properties on the variable of the model related to angular momentum distribution and quadrupoles. In the present contribution, we further explore these properties with the possibility of relating oscillatory frequencies to some high-frequency quasi-periodic oscillations models and observed data.

\end{abstract}

\maketitle

\section{Introduction}\label{intro}

Most of the black hole binaries accompanied by rapid variations in the x-ray emission on different time scales \cite{1999ApJ...520..262P,2009MNRAS.397L.101I,2011MNRAS.415.2323I,2016AN....337..398M}. The important features of the X-ray emission from these systems is quasi-periodic oscillations (QPOs). The QPOs have been detected for several black hole binaries and they are classified as low frequency QPOs ($\leq 30$ Hz) and high frequency QPOs ($\geq 60$ Hz).

There is a good agreement that strong correlations exist between the QPOs frequency and the accretion disc parameters like inner disc radius; however, the physics of their origin is not yet clear. Therefore, QPOs can provide a possibility to probe the strong gravity near compact objects. They also may be used to calculate the spin and mass of the compact object. Several models try to explain QPOs based on the geometrically thin and optically thick accretion disc, which are truncated at a radius greater than the innermost stable circular orbit.  For example, the Relativistic Precession model (RPm) assumes QPOs are produced by a local motion of accreted inhomogeneities and tries to relate the QPO frequencies to the Keplerian and periastron precession frequency \citep{1998ApJ...492L..59S,PhysRevLett.82.17}. In this respect, the properties of the orbital motion frequencies have been extensively studied \citep{1981GReGr..13..899A,1986Ap&SS.124..137A,2004PhRvD..69h4022A,2010ApJ...714..748T,2010CQGra..27d5001B,2012ApJ...760..138T,2013arXiv1309.6396G,2014arXiv1411.4811M,2015CQGra..32p5009K,2016EPJC...76...32S,2016EPJC...76..414A,2019PhRvD.100h4038T,2020Galax...8...76A,2020PhyS...95h5008Y} among many others. 

During these years, this model modified in different ways. For example, a different class presented the collective motion of accreted matter considering normal modes of thin accretion disk oscillations, also slender and non-slender tori oscillations e.g. \citep{2001ApJ...559L..25W,2001PASJ...53....1K,2003MNRAS.344L..37R,2012ApJ...752L..18W,2013A&A...552A..10S,2016MNRAS.457L..19T,2018MNRAS.474.3967D,2020MNRAS.497.1066M,2020PASJ...72...38K,2020A&A...643A..31K}. In particular, oscillations of geometrically thick discs are notable since this disk can radiate significantly and they reveal a long-term oscillatory behaviour with the duration of tens of orbital periods. The oscillation modes of geometrically thick discs are studied through the vertically averaging technique and local perturbative analyses in \cite{10.1046/j.1365-8711.2003.06474.x,10.1046/j.1365-8711.2003.07023.x,2004MNRAS.354.1040M}. In these works, the expression of the radial epicyclic frequency is derived for a fluid disk in these works. In fact, for interpreting the observed QPOs frequencies, maybe one important consideration is to pay attention to using the expression of the radial epicyclic frequency for a test particle. Therefore, one further step can be the investigation of the radial epicyclic frequency in a perfect fluid disk. This paper aims to extend this approach with different angular momentum distributions to the background of a distorted deformed compact object due to quadrupoles. We studied the QPOs with the test particle approach in this background in \cite{2022MNRAS.513.3399F}. Indeed, in the astrophysics area, people attempt to determine the observable predictions of strong-field images of accretion flows in many ways. In this respect, this approach may provide an opportunity to take quadrupole moments as the additional physical degrees of freedom to the central compact object and its surroundings. Moreover, there is no doubt about the fundamental importance of gravitational waves in physics, where the experimental evidence finally supported the purely theoretical research in this area \cite{PhysRevLett.116.061102}. In fact, in the case of extreme mass ratio inspiral (EMRI), one can extract the multipole moments from the gravitational wave signal, and any non-Kerr multipole moments should be encoded in the waves \cite{1995PhRvD..52.5707R}. A study of gravitational waves generated in an EMRI is in progress. This space-time is briefly described in Section \ref{spacetime}. In addition, we examine the results of this study by using the data of three high-frequency QPOs observed in the microquasars GRS 1915+105, XTE 1550-564 and GRO 1655-40. 

The plan of the paper is as follows. In Section \ref{purt} we briefly explain the relativistic tori model and the local perturbation approach. In Section \ref{spacetime} the desired background is explained. Results and discussion is presented in Section \ref{result}. Finally, the conclusion is summarized in Section \ref{summary}. In this work, we use the metric signature as $(-,+,+,+)$ and geometrized unit system $G=1=c$ (However, for an astrophysical application, we will use SI units). Latin indices run from 1 to 3, while Greek ones take values from $0$ to $3$.

\section{Perturbation of relativistic tori} \label{purt}

The relativistic tori in a stationary and axisymmetric space-time, is the configuration of the perfect fluid described by polytropic equation of state of this type $p=k \rho^{\gamma}$, with the polytropic constant $k$ and the adiabatic index $\gamma=\frac{d\ln p}{d \ln \rho}$. This equation implies the conservation of entropy for a perfect fluid. The fluid is determined by its rotation in the azimuthal direction with the four-velocity, specific angular momentum and the angular velocity 
\begin{align}
    &u^{\mu}=(u^t,0,0,u^{\phi}),\\
    &l=-\frac{u_{\phi}}{u_t},\\
    &\Omega=\frac{u^{\phi}}{u^t}.\
\end{align}
The shape of the tori with rest-mass density $\rho$ is determined by equi-pressure surfaces profile which specify by
the basic conservation equations 

\begin{align}
    &\nabla_{\beta}(\rho u^{\beta}) =0,\\
    &\nabla_{\beta}T^{\beta \alpha} =0.\
\end{align}
where $\nabla$ is the covariant derivative, and $T^{\beta \alpha}$ is the stress-energy tensor
\begin{align}
    T^{\beta \alpha}=&w u^{\beta}u^{\alpha}+pg^{\beta \alpha},\
\end{align}
where $\omega$ is the enthalpy and $p$ is the pressure.
From these relation the Bernoulli-type equations follows for pressure distribution \cite{1978A&A....63..221A} 

\begin{align}\label{A7}
\frac{1}{w}\nabla_{i}p=-\nabla_{i}\ln u_{t}+\frac{\Omega\nabla{}_{i}\ell}{1-\Omega\ell}. 
\end{align}
In order to integrate and solve this equation analytically, we need to consider $\Omega=\Omega(l)$ and determine distribution of angular momentum. Of course, the simplest choice for $l$ to have a finite disc size is being a constant, say $l_{\rm cons.} $ chosen between the specific angular momentum at the marginally stable orbit $l_{\rm ms}$ and the specific angular momentum at the marginally bound orbit $l_{\rm mb}$. However, in what follows we also consider non-constant angular momenum. In addition, there are two important locations in this model; namely the cusp, $r_{\rm cusp}$ and the center $r_{\rm c}$. They are defined as the places when the Keplerian specific angular momentum is equal to $l_{\rm cons.}$. The center is defined as where we have the maximum density in the torus. The inner and outer edges of the tori can be specify by asking they lie on the same equi-potential surface. In this work, we also worked with a non-constant angular momentum distribution in which it has two important properties features of the angular momentum distribution in accretion disks. This distribution on the equatorial plane and far from the object is slightly sub-Keplerian, but closer it becomes slightly super-Keplerian and in the plunging region, it becomes again sub-Keplerian and almost constant \cite{2009A&A...498..471Q}

\begin{align} \label{osyqianangular}
 \ell(r,\theta)=
   \left\{
  \begin{array}{@{}ll@{}}
  \ell_0\left(\frac{\ell_{\rm K(r)}}{\ell_0}\right)^{\gamma}\sin^{2\delta}, & r\geq r_{\rm ms}, \\
     \ell_0(\zeta)^{-\gamma}\sin^{2\delta}, & r<r_{\rm ms},
    \end{array}\right.
\end{align}
where $\ell_0=\zeta \ell_{\rm K}(r_{\rm ms})$, and $\ell_{\rm K}$ is the Keplerian angular momentum in the equatorial plane, and

\begin{align}
   0\leq \gamma\leq 1, \quad -1\leq \delta \leq 1, \quad -1\leq \zeta\leq \frac{\ell_{\rm K}(x_{\rm mb})}{\ell_{\rm K}(x_{\rm ms})}.
\end{align}
where $\ell_{\rm ms}(r)$ is calculated on the equatorial plane via considering $\Omega_{\rm ms}$. In the following subsection, we introduce the local perturbations in this background following the procedure in \cite{10.1046/j.1365-8711.2003.07023.x,2004MNRAS.354.1040M}.

\subsection{Local perturbation} \label{local}

In this method, to reduce one spatial dimension for simplicity to solve analytically, the vertically integrated and vertically averaged technique along the direction perpendicular to the equatorial plane is assumed for quantities. Beside, the local approach leads to derive the local dispersion relation for the oscillations
in relativistic non-Keplerian discs in a given background.

In the following we discuss briefly the local perturbation approach for the circular orbiting test particle and a perfect fluid. However, instead of power-law angular momentum distribution used in \cite{10.1046/j.1365-8711.2003.07023.x,2004MNRAS.354.1040M}, here we used the one that mentioned earlier via the equation \eqref{osyqianangular}.

In this approach, we are interested in the vicinity of the equatorial plane $|y-0|\ll 1$, which is different from consider just the equatorial plane of course since here use the vertically integrated quantities corresponding to collapsing the vertical structure of the tori on the equatorial plane and not just consider the equatorial slice of the tori and calculate the quantities in the equatorial plane. This approach is very similar to the technique used for the Thin accretion disk \cite{1973A&A....24..337S,1973blho.conf..343N}. 

Following \cite{10.1046/j.1365-8711.2003.07023.x} we introduce the vertically integrated pressure $P$, rest-mass density $\Sigma$, velocity components $U(r)$ and $W(r)$ in terms of the local thickness $H=H(r)$ of tori, respectively as

\begin{align}
P(r)&:=\int_{-H}^{H} p dz,\\
\Sigma(r)&:=\int_{-H}^{H} \omega dz,\\
U(r)&:=\frac{1}{2H}\int_{-H}^{H} v^r dz,\\
W(r)&:=\frac{1}{2H}\int_{-H}^{H} v^{\phi} dz,\
\end{align}
where $v^i$ are the three-velocity of the fluid. We also need to modify the equation of state in a consistent way. To do so we define the adiabatic index $\Gamma=\frac{d\ln P}{d \ln \Sigma}$ so that

\begin{align}\label{effeof}
    P=\mathcal{K}\Sigma^{\Gamma},
\end{align}
where $\mathcal{K}$ is the corresponding polytropic constant. However, equation \eqref{effeof} does not represent a vertically integrated polytropic equation of state if $\Gamma\neq 1$. In addition, to get ride of the height dependence of $\Gamma=\Gamma(r,z)$ for simplicity one can assume $p$ and $\rho$ have a weak dependence on height, so that they can be stated in terms of their values at the equatorial plane. Therefore, we can use this modified expression for the equation of state

\begin{align}
    P=k\Sigma^{\gamma}.
\end{align}
With these quantities the dynamics of the torus is fully determined \cite{1992pavi.book.....S,10.1046/j.1365-8711.2003.07023.x} once we choose the parameters of the model. Next, we introduce harmonic Eulerian perturbations 

\begin{align} \label{pertosc}
    \begin{pmatrix}
  {\delta U} \\
  {\delta W} \\
   {\delta Q} \\
  \end{pmatrix}
  \sim e^{-i\sigma t +ikr}
\end{align}
where for the the fluid pressure we consider

\begin{align}
  \delta Q= \frac{\delta P}{\Sigma}.  
\end{align}
In fact, the harmonic spatial dependence in this relations as
the signature of the local perturbation is valid for when the wavelength of the perturbations is smaller than the radial variations in the equilibrium configuration, the Wenzel–Kramers–Brillouin (WKB)
approximation. By applying the above perturbations in the equilibrium tori model and  considering only up to the first-order terms and neglecting the time derivatives, we derive the perturbation equations \cite{10.1046/j.1365-8711.2003.07023.x}. This linear system of differential equations has a non-trivial solution by searching for the zeros of determinants of coefficients matrix, which simply leads to the dispersion relation 

\begin{align} \label{dispertionrel}
   \sigma^2=\kappa^2 + f(r) k^2c_s^2,
\end{align}
where $k_r$ is the radial epicyclic frequency, and $c_s$ is the local sound velocity. The function $f(r)$ in this relation changes for different backgrounds. In Newtonian set-up $f(r)=1$ and has different expression in Schwarzschild space-time \cite[equation 42]{10.1046/j.1365-8711.2003.07023.x}, or in the Kerr space-time \cite[equation 17]{2004MNRAS.354.1040M}. Moreover, it is assumed that the perturbation wavelength is much smaller than the radius of the disk, and the imaginary part of the epicyclic frequency is also neglected \cite{10.1046/j.1365-8711.2003.07023.x}.

In fact, the first term in the dispersion relation is the radial oscillation of a fluid element when due to a restoring centrifugal force, it is infinitesimally displaced from its equilibrium with no change in the angular momentum. This produces inertial oscillations with the frequency $\kappa$. However, there exist also a vertical epicyclic frequency with an amplitude much smaller than the radial one \cite{2004ApJ...603L..89K}, therefore we neglected here. The second term, the frequency $kc_s$ is due to pressure gradients in compressible fluids that leads to the acoustic oscillations \footnote{Both of these terms are collectively referred to as inertial–acoustic waves.}.

To study the radial epicyclic oscillation in this approach, one can consider different angular momentum distributions. However, for the Keplerian angular momentum distribution where $\Omega \propto r^{-\frac{3}{2}}$ the radial epicyclic frequency is equal to the orbital frequency $\kappa^2_r = \Omega^2$. Besides, the radial epicyclic frequency vanishes for the constant specific angular momentum dictating by the corresponding equations \cite{10.1046/j.1365-8711.2003.07023.x,2004MNRAS.354.1040M}. Thus, for this investigation we used non-constant angular momentum distribution presented in \cite{2009A&A...498..471Q}. In the next section, the background space-time that we are interested to consider in this work is briefly explained.

\section{Space-time} \label{spacetime}
In many years, the Weyl's family of solutions  \cite{doi:10.1002/andp.19173591804} to the static vacuum Einstein equations, have been used to modeling the exterior gravitational field of compact axially symmetric bodies. In this regard, the simplest and applicable generalization of the Schwarzschild family is the $\rm q$-metric. It describes static, axially symmetric, and asymptotically flat solutions with quadrupole moment describing the deviation from the spherical body. The metric represents the exterior gravitational field of an isolated compact object \cite{doi:10.1063/1.1705005,PhysRevD.2.2119,Quevedo:2010vx,2011IJMPD..20.1779Q}. This metric has been further generalized by relaxing the assumption of isolated compact object via considering the exterior distribution of mass in its vicinity. It is similar to considering additional external gravitational field, like adding a magnetic surrounding \cite{1976JMP....17...54E}. This generalization is also characterising by quadrupole \cite{universe8030195}. 
In fact, the presence of a quadrupole, can change the geometric properties of space-time drastically. The metric in
the prolate spheroidal coordinates is presented as follows  \cite{universe8030195}


\begin{align}\label{EImetric}
	d s^2 &= - \left( \frac{x-1}{x+1} \right)^{(1+\alpha)} e^{2\hat{\psi}} d t^2+ M^2(x^2-1) e^{-2\hat{\psi}} \nonumber\\
	 &\left( \frac{x+1}{x-1} \right)^{(1+\alpha)}\left[ \left(\frac{x^2-1}{x^2-y^2}\right)^{\alpha(2+\alpha)}e^{2\hat{\gamma}}\right.\nonumber\\
	 &\left. \left( \frac{d x^2}{x^2-1}+\frac{d y^2}{1-y^2} \right)+(1-y^2) d{\phi}^2\right],\
\end{align}
where $t \in (-\infty, +\infty)$, $x \in (1, +\infty)$, $y \in [-1,1]$, and $\phi \in [0, 2\pi)$. Also $\alpha \in (-1,\infty)$ is the quadrupole associated with the central compact object. Nevertheless, we consider relatively small quadrupole moment; therefore, a physically interior solution can be fit to this external solution \cite{Quevedo:2010vx}. quadrupole $\alpha$ determines the deviation from spherical shape and how the mass distribution is stretched-out along some axis. If $\alpha>0$ we have an oblate object and for $\alpha<0$ we have a prolate shape. If $\alpha=0$, $\hat{\psi}=0$, $\hat{\gamma}=0$ the Schwarzschild metric is recovered. Up to the quadrupole the external field terms read as follows

\begin{align} \label{1111}
\hat{\psi} & = -\frac{\beta}{2}\left[-3x^2y^2+x^2+y^2-1\right],\nonumber\\
\hat{\gamma} & = -2x\beta(1+\alpha)(1-y^2)\nonumber\\
  &+\frac{\beta^2}{4}(x^2-1)(1-y^2)(-9x^2y^2+x^2+y^2-1),\
  \end{align}
where $\beta$ is the quadrupole associated with the external fields. To have some intuitive picture about the role of $\beta$ we can consider Newtonian gravity (we show quadrupole in Newtonian picture by $\beta_N$). In Newtonian theory it is rather familiar piece of knowledge that the multipole expansion dominated by a quadrupole moment $\beta_N$ can be modelled by two equal point-like masses $m$ located on some axis, say $z$, at some distance from the center, also an infinitesimally thin ring of the mass $M$ and radius $R$ located at the plane perpendicular to this axis. If the contribution of the point-like masses to the gravitational field is greater than  the ring, then $\beta_N<0$, and if for ring is greater we have $\beta_N>0$. If $\beta_N<0$, then there exist a net force  directed toward the $z$-axis. This force creates a potential barrier. If $\beta_N>0$, there is a net force directed to the ring, outward from the central object. This force balances the gravitational of the central source and the external fields.

The relation between this coordinates system and the Schwarzschild like coordinates is given by

\begin{align}\label{transf1}
 x =\frac{r}{M}-1 \,, \quad  y= \cos\theta\,.
\end{align}
In the following we present the relativistic tori in this background and analysis the epicyclic frequencies as mentioned earlier. 


\begin{figure*}
    \centering
    \begin{tabular}{cc} \includegraphics[width=0.3\hsize]{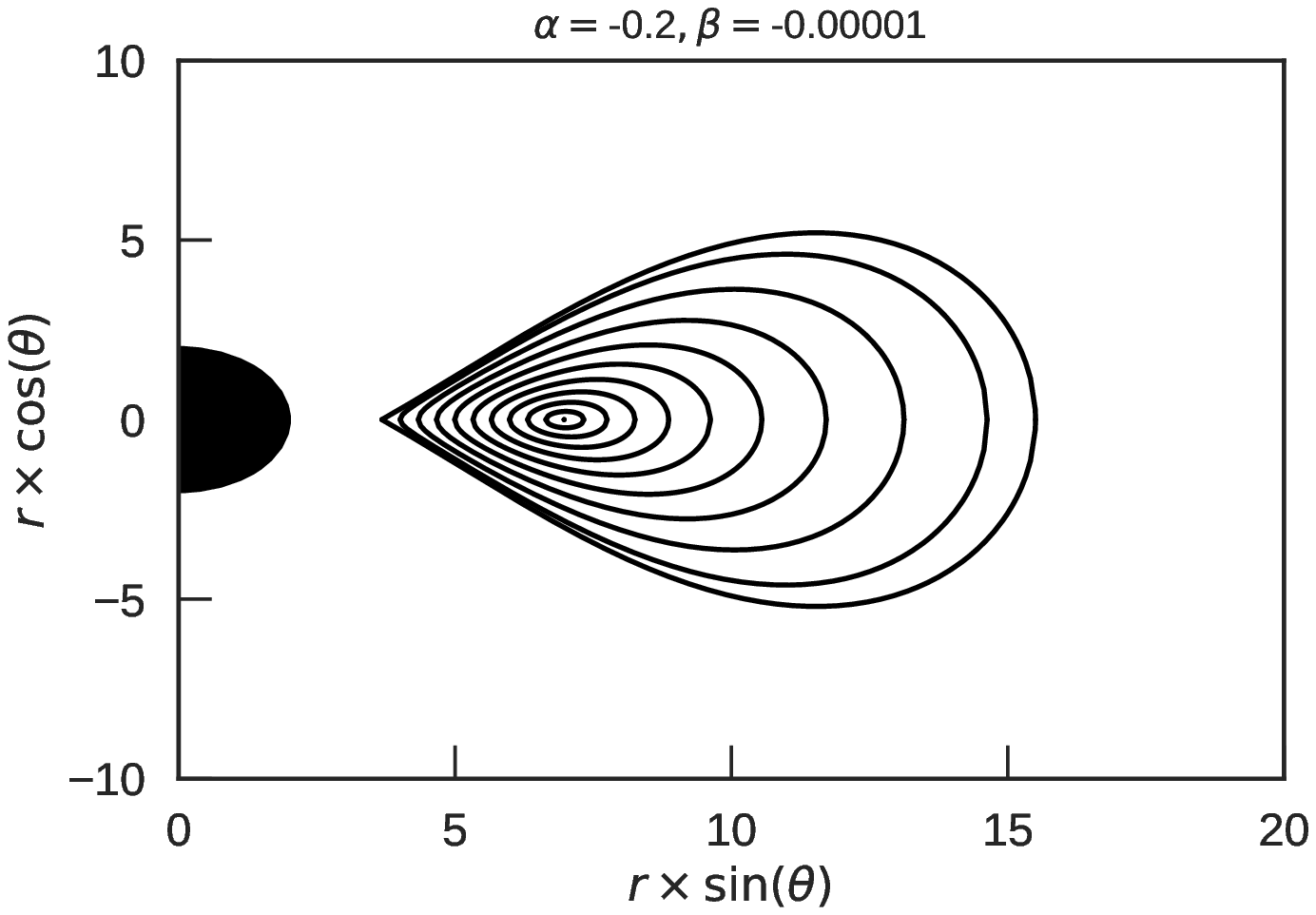}&
      \includegraphics[width=0.3\hsize]{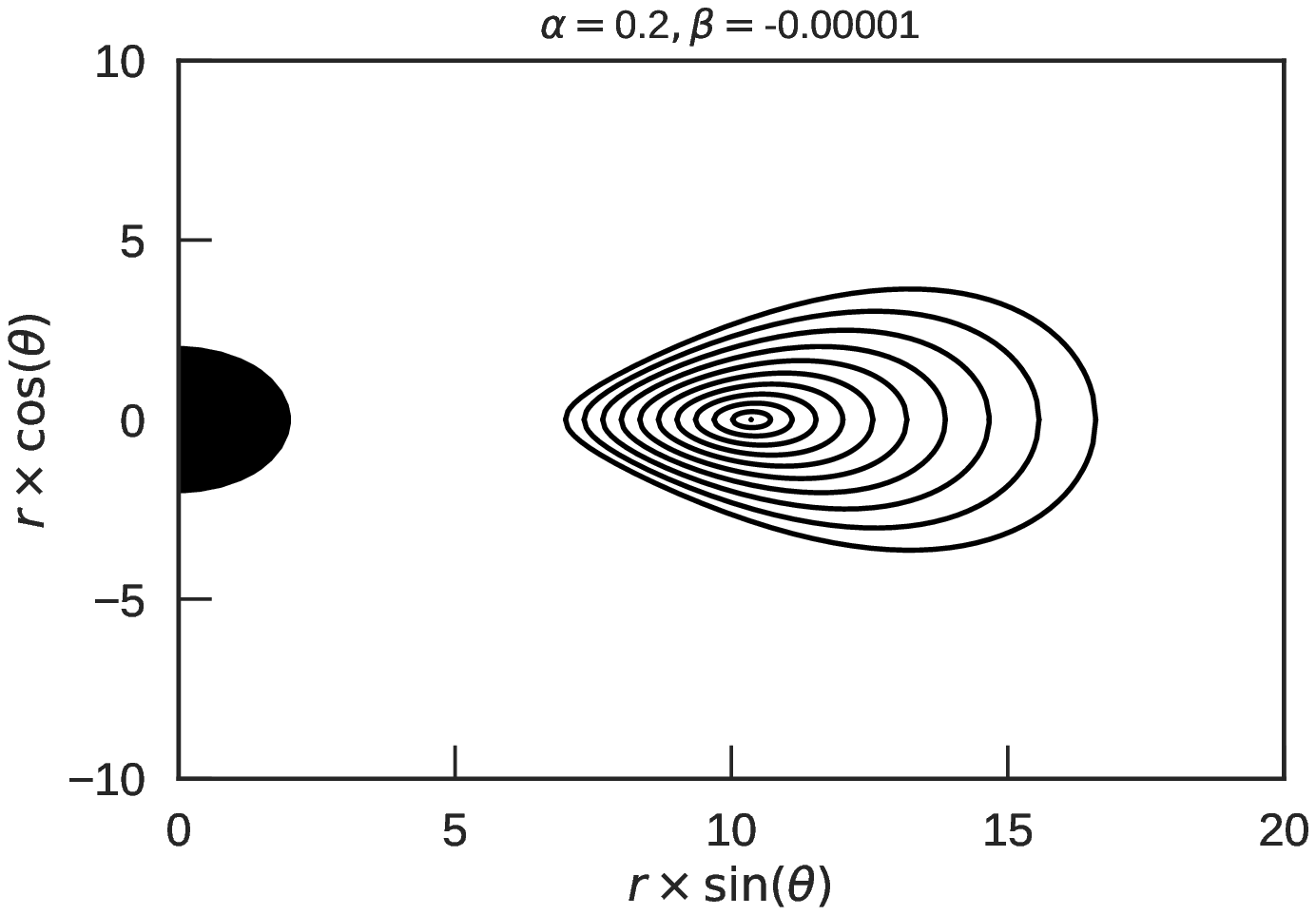}\\
     \includegraphics[width=0.3\hsize]{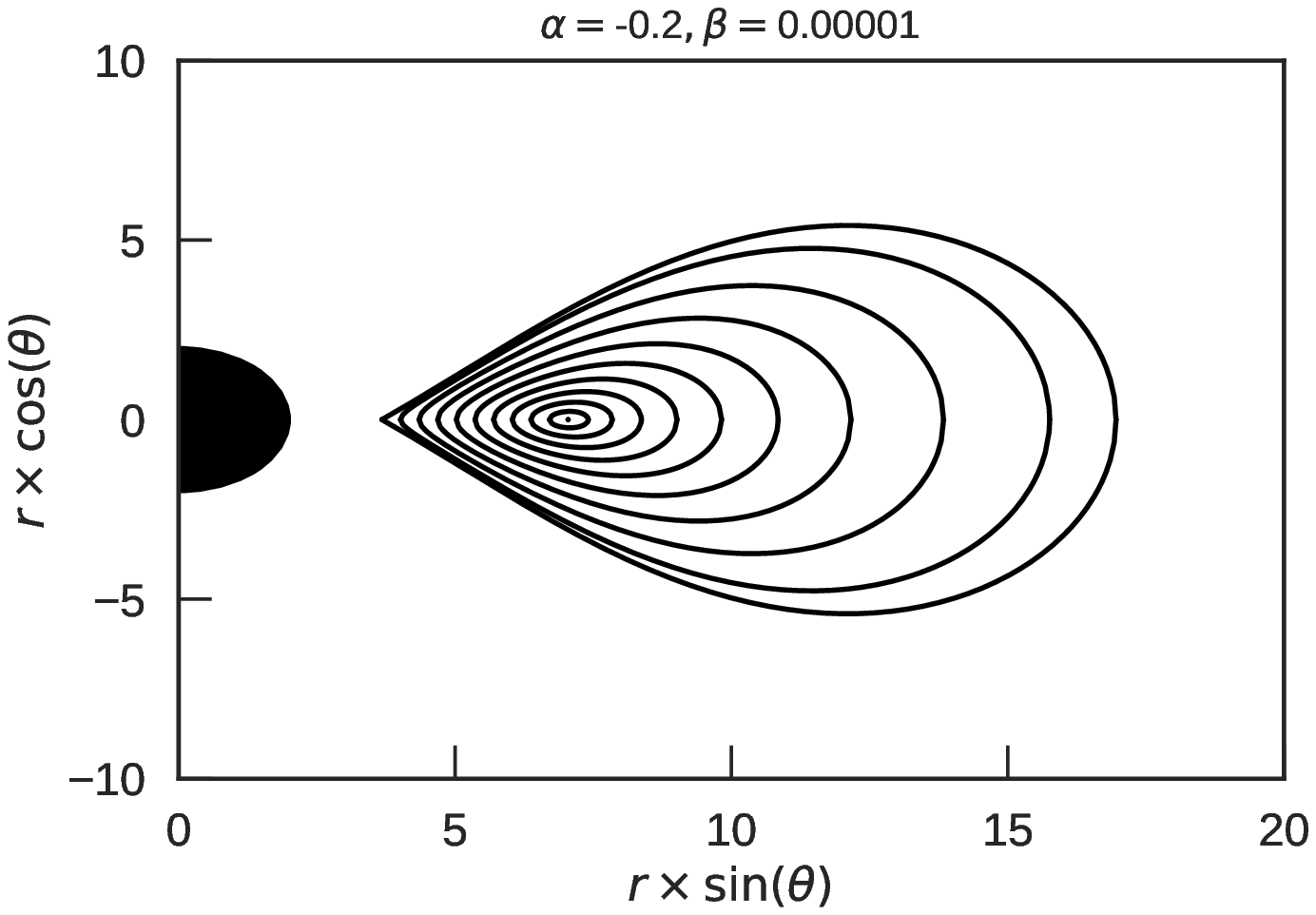}&
     \includegraphics[width=0.3\hsize]{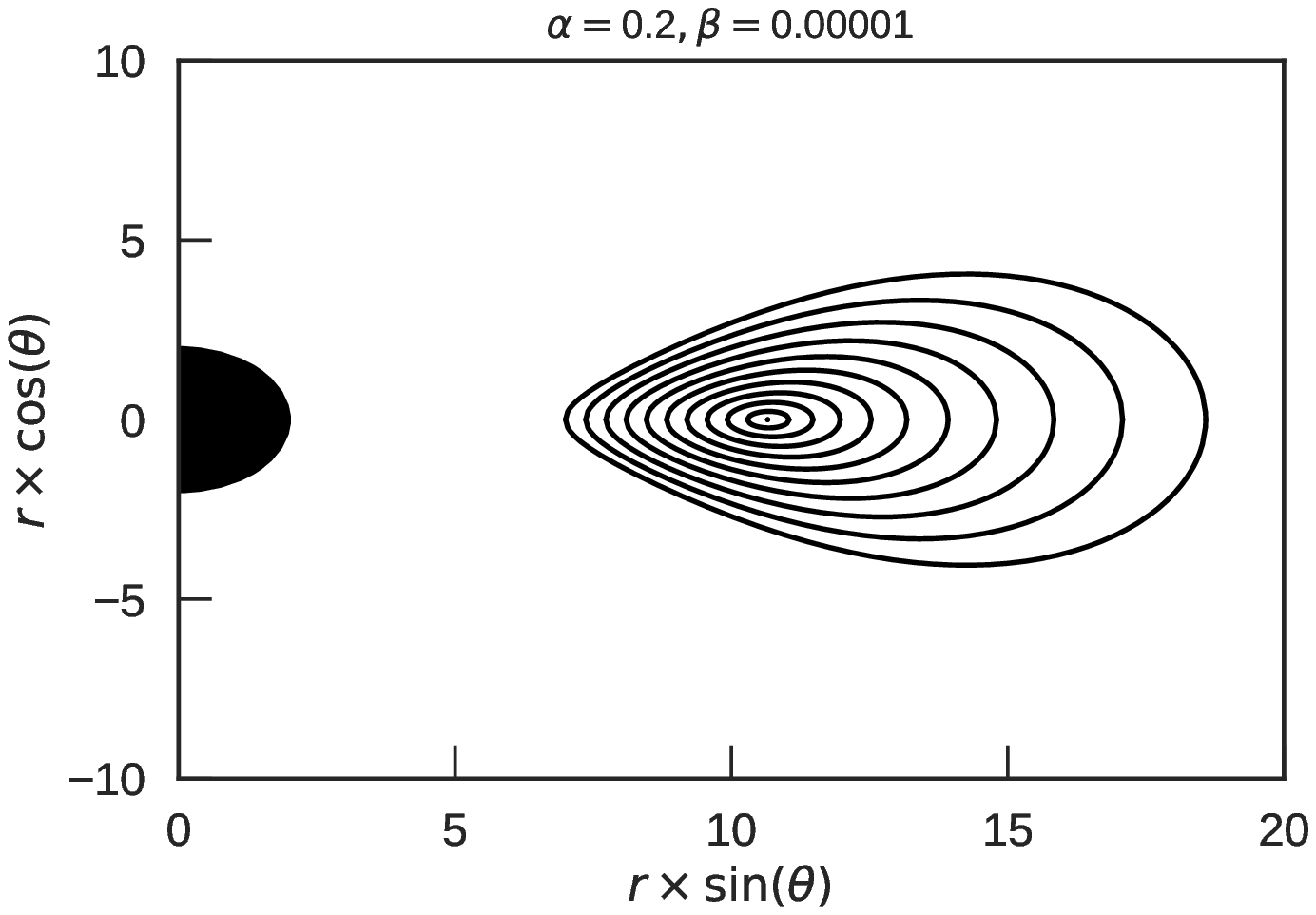}
     \end{tabular}  
    \caption{Equipotential surfaces for tori with a constant angular momentum distribution $\ell_{\rm cons.}=(\ell_{\rm ms}+\ell_{\rm mb})/2$ for different values of  of $\alpha$ and $\beta$.}
    \label{fig:MapConstpq}
\end{figure*}

\section{Results and discussion} \label{result}

Before start studying the epicyclic frequency we investigate the model in this background briefly.

\subsection{Relativistic tori}

The equation \eqref{A7} for radial component in this background obtained as

\begin{figure*}
    \centering
    \begin{tabular}{ccc}

    \includegraphics[width=0.3\hsize]{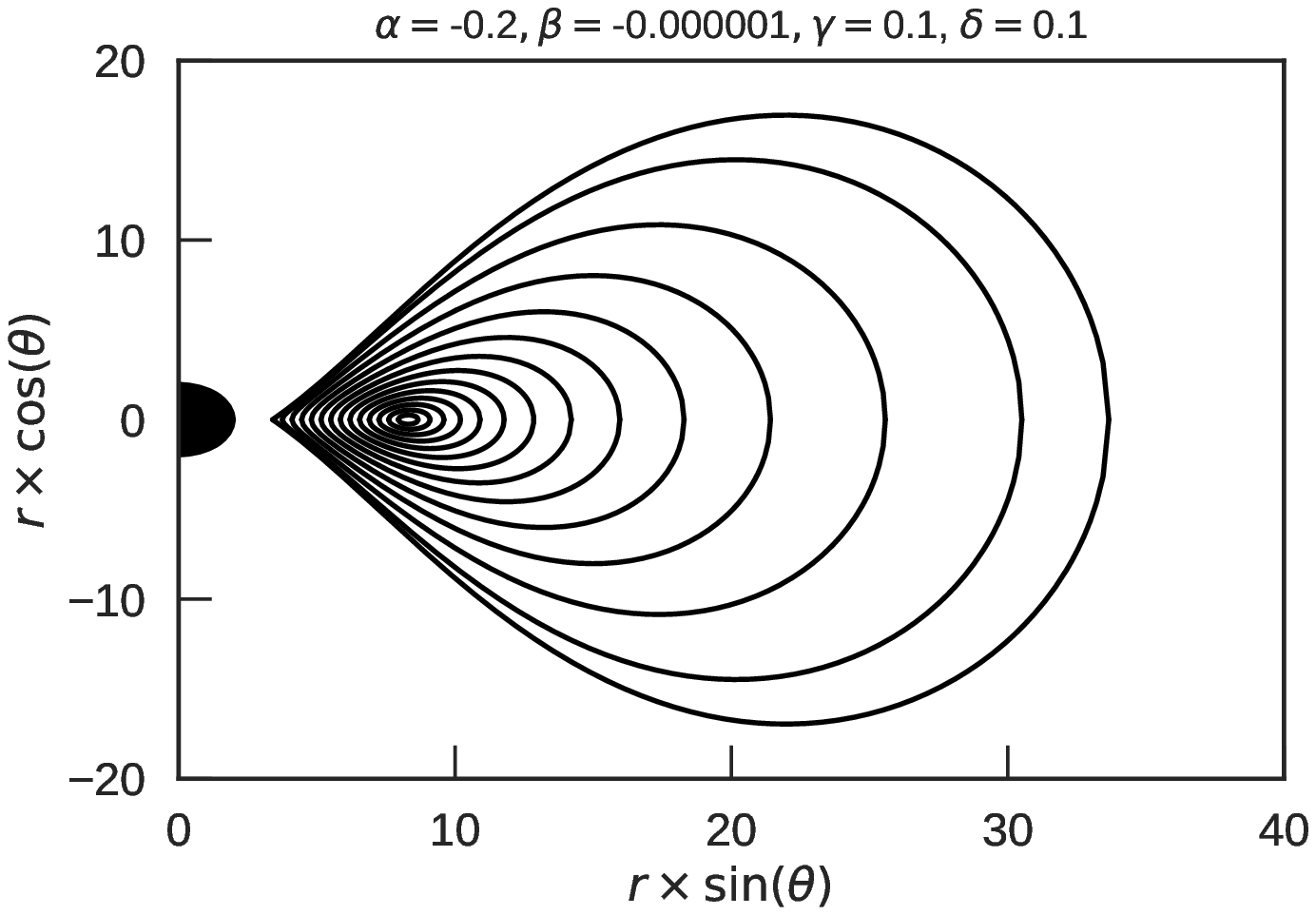} &
     \includegraphics[width=0.3\hsize]{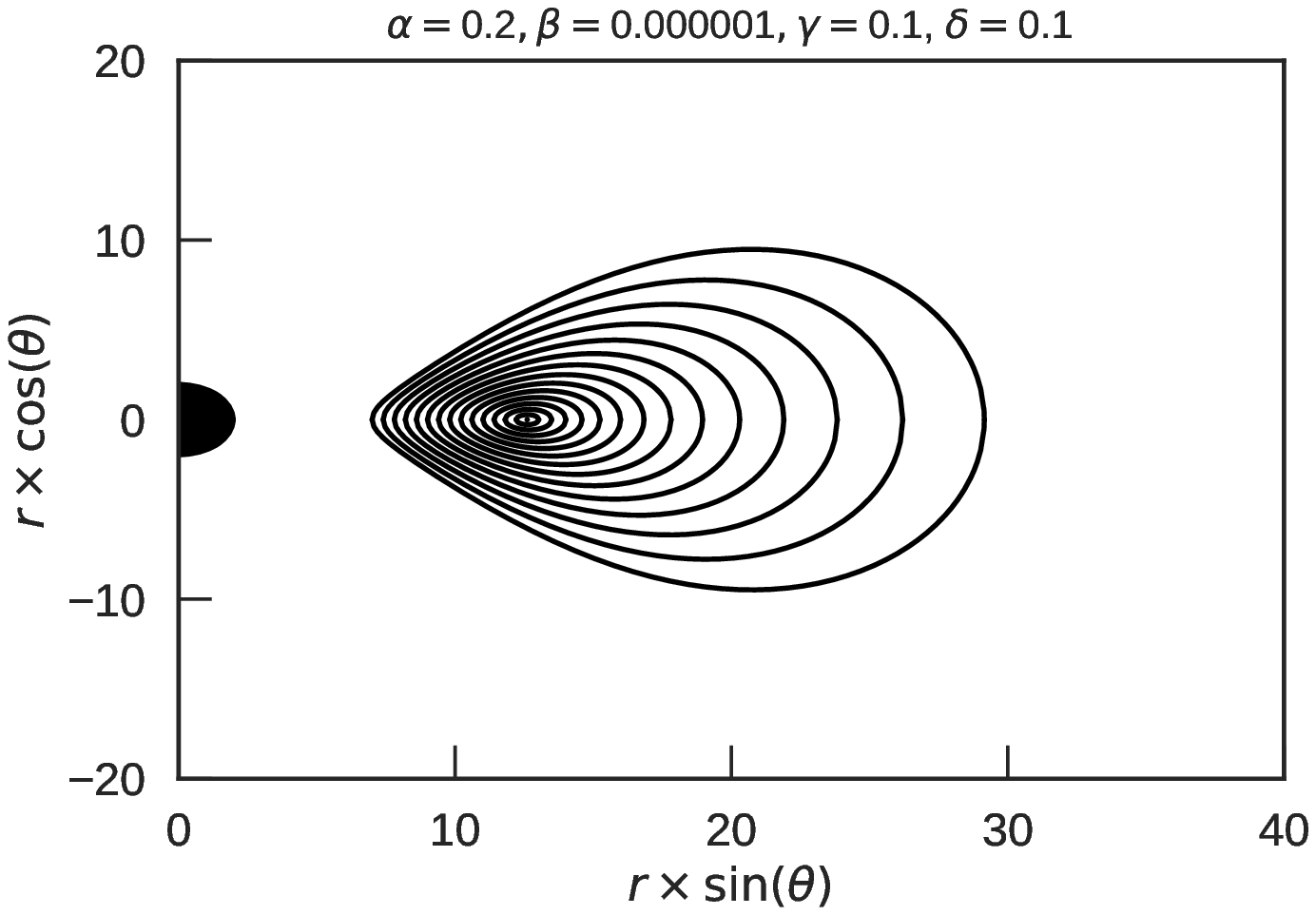}\\
     \includegraphics[width=0.3\hsize]{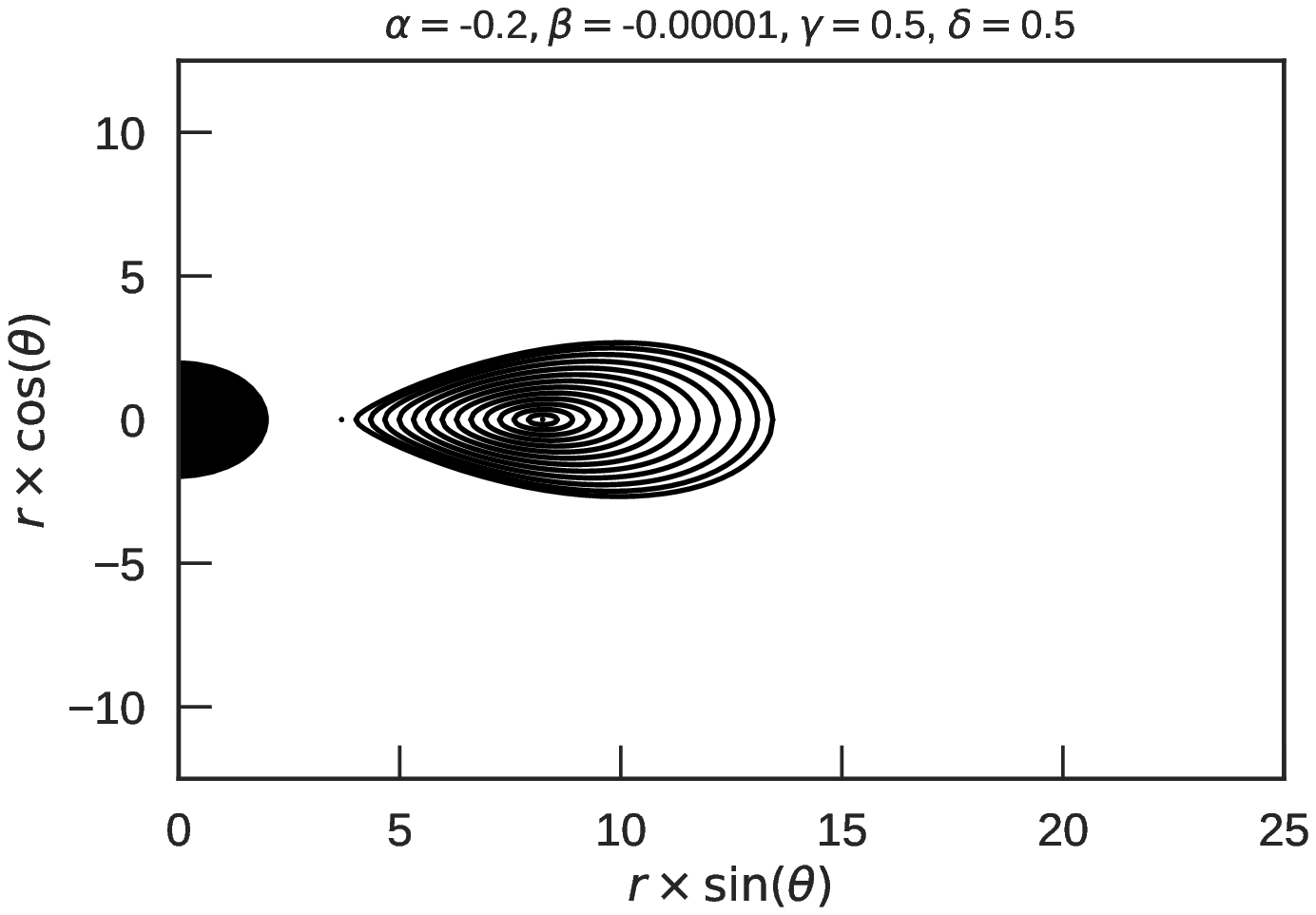} &
     \includegraphics[width=0.3\hsize]{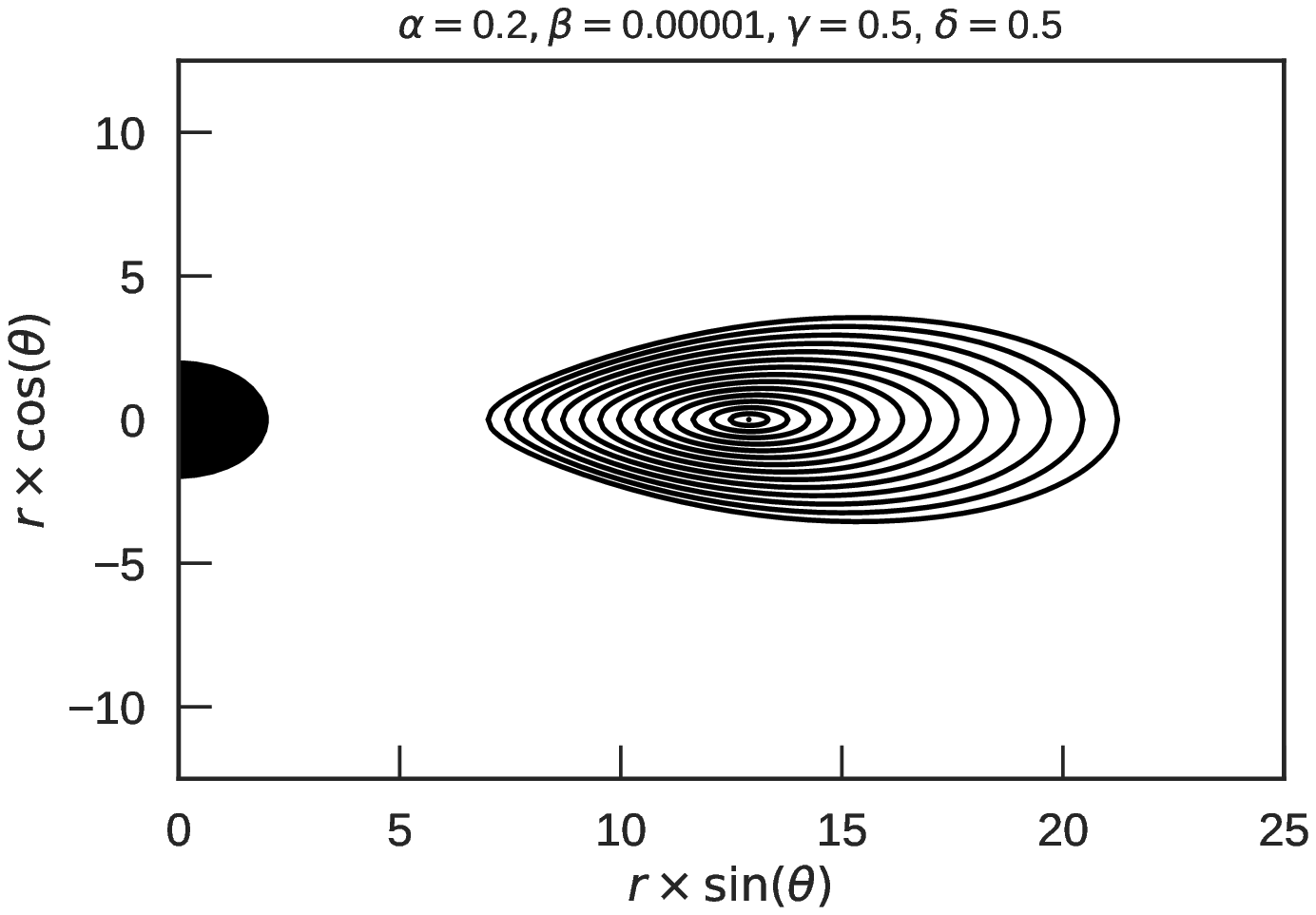}\\
     \end{tabular}  
    \caption{Equipotential surfaces for tori with the non-constant angular momentum distribution for various values of combination of $\alpha$ and $\beta$.}
    \label{fig:MapQianpq}
\end{figure*}

\begin{widetext}
\begin{equation}
\frac{1}{\omega}\partial_r p=\frac{r\left(\frac{r}{r-2}\right)^{\alpha}(\alpha-(r-2)(1+\beta(r-1) r)) \Omega^{2}+e^{-\beta(r-2) r+2\hat{\psi}}(r-2)^{1+\alpha} r^{-2-\alpha}(1+\alpha+(r-2) r \partial_r \hat{\psi} )}{(r-2)\left(-e^{-\beta(r-2) r+2 \hat{\psi}}\left(\frac{r-2}{r}\right)^{1+\alpha}+r^{2}\left(\frac{r}{r-2}\right)^{\alpha} \Omega^{2}\right)}.
\end{equation}
\end{widetext}
As it pointed out, to be able to integrate and solve this equation one needs to specify the angular momentum distribution. In this work, first we consider constant distribution to see the effects of quadrupoles in the metric, and then used the distribution in the equation \eqref{osyqianangular}. Figure \ref{fig:MapConstpq} shows the disk configuration for $\ell_{\rm cons.}=(\ell_{\rm ms}+\ell_{\rm mb})/2$, where $\l_{\rm mb}$ and $\l_{\rm ms}$ are calculated for chosen values of quadrupoles written on the plots. In this Figure we can see the influence of quadrupoles on the shape and size of the tori. We see that, in general, for negative $\alpha$ which is correspond to prolate object we have larger size disk rather than Schwarzschild and positive ones. For $\alpha>0$ the disk is smaller and is formed far from the central object comparing to negative $\alpha$. On the other hand, having positive quadrupole in the external field $\beta>0$ causes to have a bigger disk structure and more extended radially. However, it is worth mentioning that the overall shape of disk is more influenced by $\alpha$ which is the quadrupole links to the central object. In Figure \ref{fig:MapQianpq} we see the disk configuration in the present of different angular momentum. In this Figure, we used the same values for quadrupoles as in Figure \ref{fig:MapConstpq} to see the effect of angular momentum distribution clearly. We see that choosing the angular momentum profile has a strong effect on the shape and size of the disk. The parameters of this distribution have a strong correlation with each other. However, as we chose higher values for this pairs  $(\gamma, \delta)$ the disc configuration become smaller.

In the following we study the epicyclic radial frequency. As discussed earlier for a constant angular momentum profile $\kappa=0$. Then we adopt the equation \eqref{osyqianangular} for angular momentum distribution in the rest of this work.

\subsection{Radial epicyclic frequency}

\begin{figure*}
    \centering
    \begin{tabular}{ccc}
    \includegraphics[width=0.33\hsize]{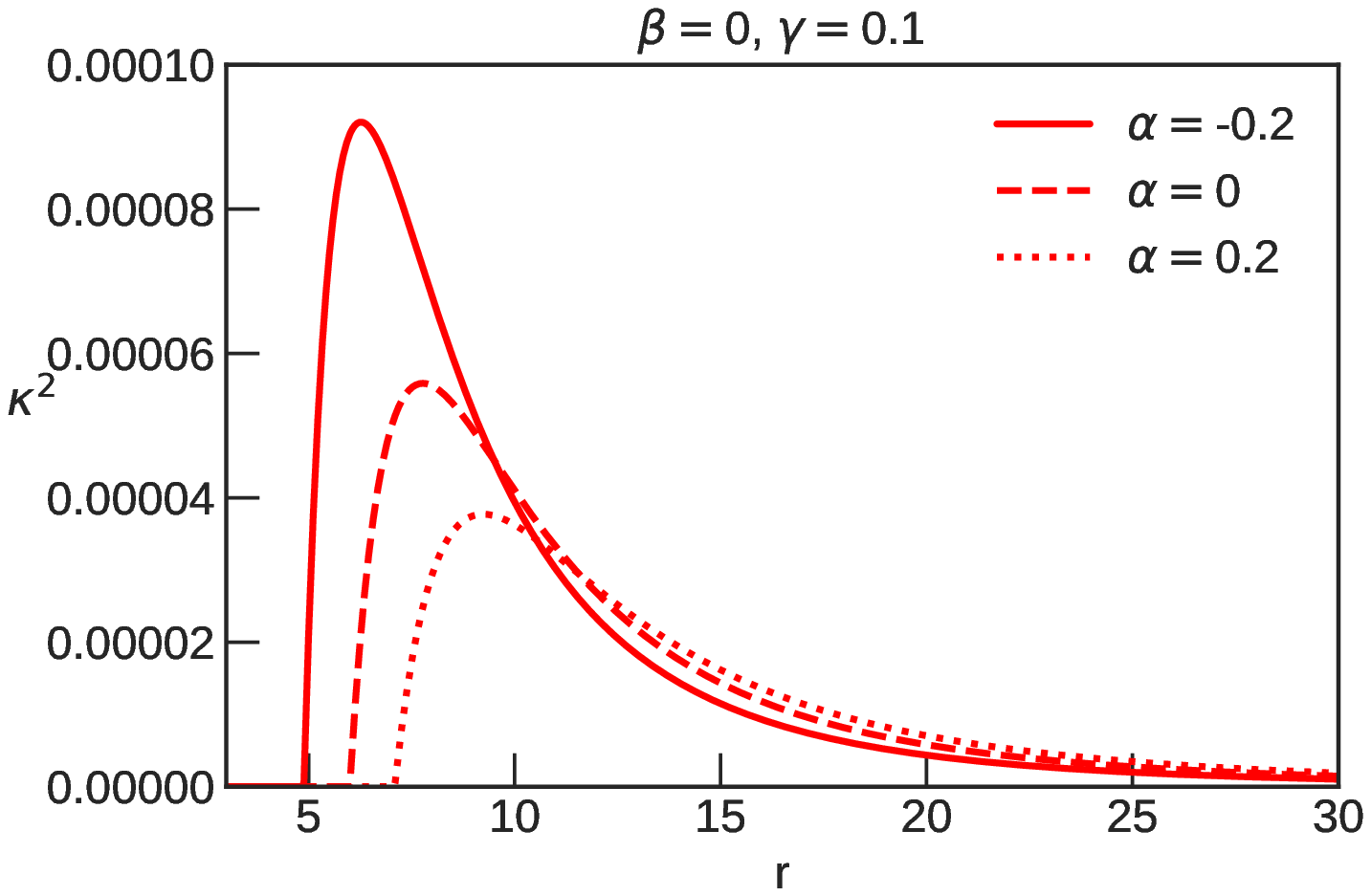}&
     \includegraphics[width=0.335\hsize]{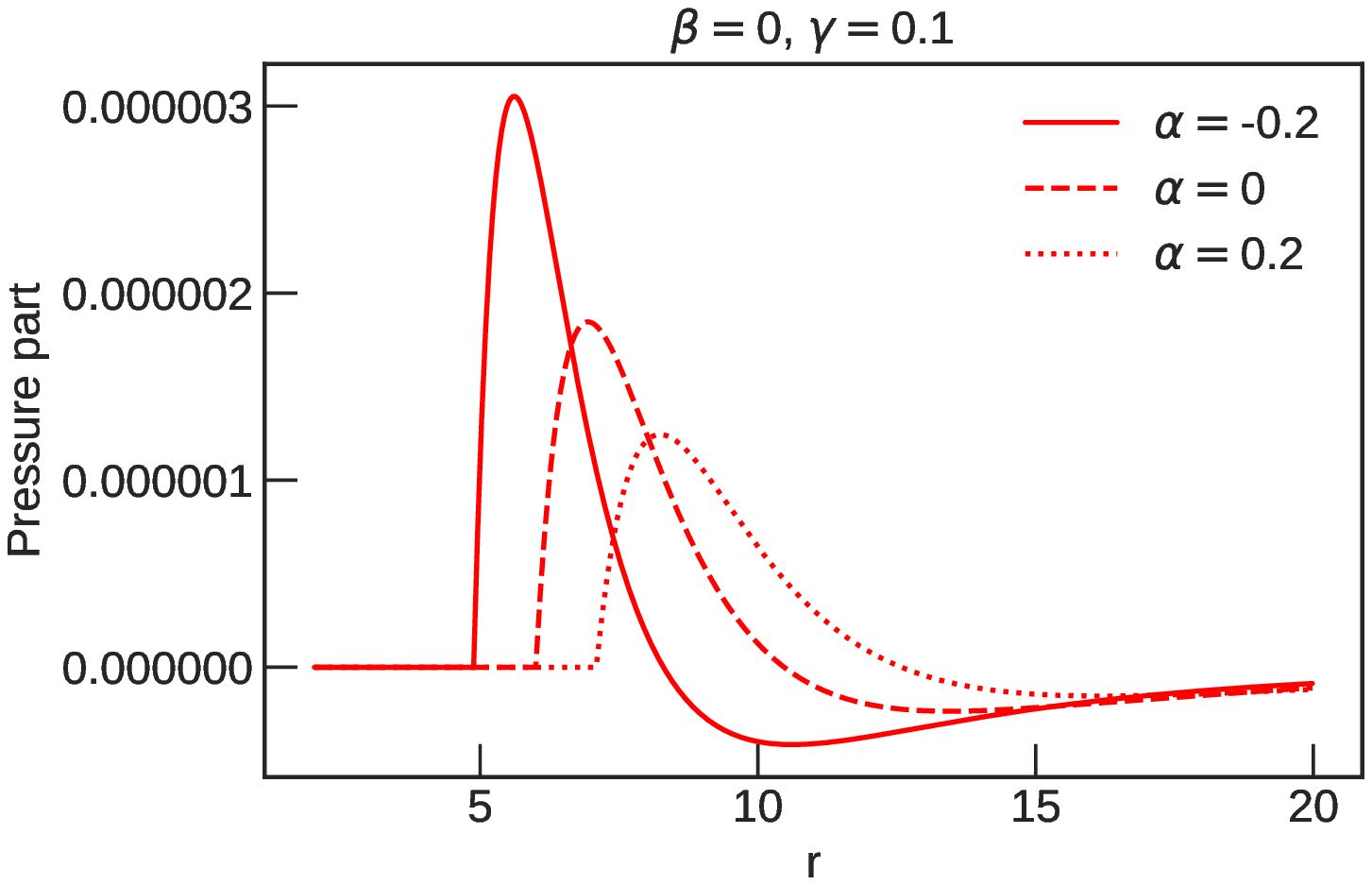} & \includegraphics[width=0.33\hsize]{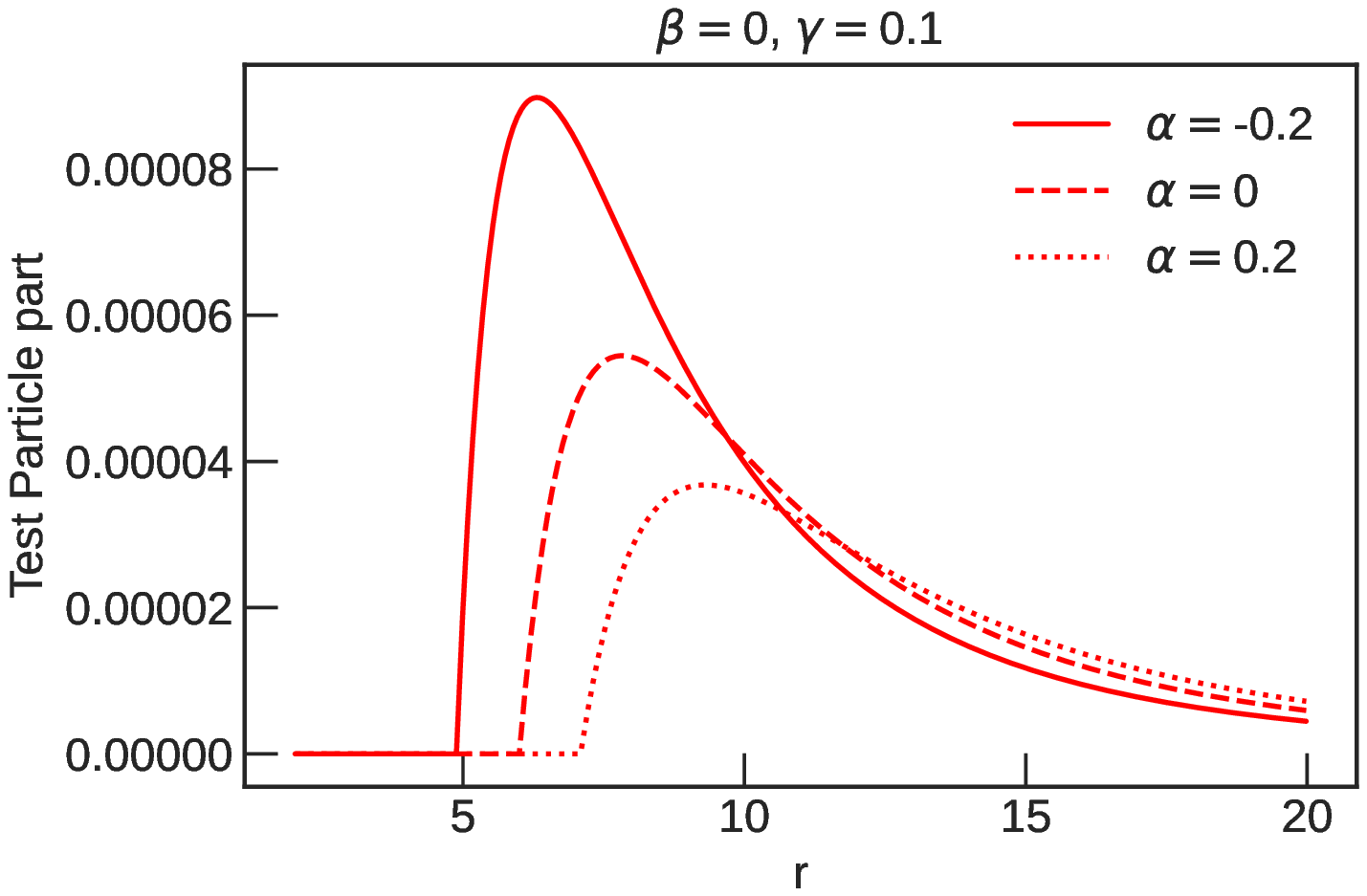}
    \end{tabular}  
    \caption{The plots present the radial epicyclic frequency squared with respect to $r$  with $\gamma=0.1$ in the angular momentum distribution \eqref{osyqianangular} for different values of $\alpha$. The right one shows $\kappa_t^2$ for a test particle (equation \ref{eqboth}). The middle one presents $\kappa_p^2$ corresponds to the pressure part, and the left one the $\kappa^2$ for the fluid.}
    \label{fig:QianVSKep1}
\end{figure*}

\begin{figure*}
    \centering
    \begin{tabular}{c}
    \includegraphics[width=0.37\hsize]{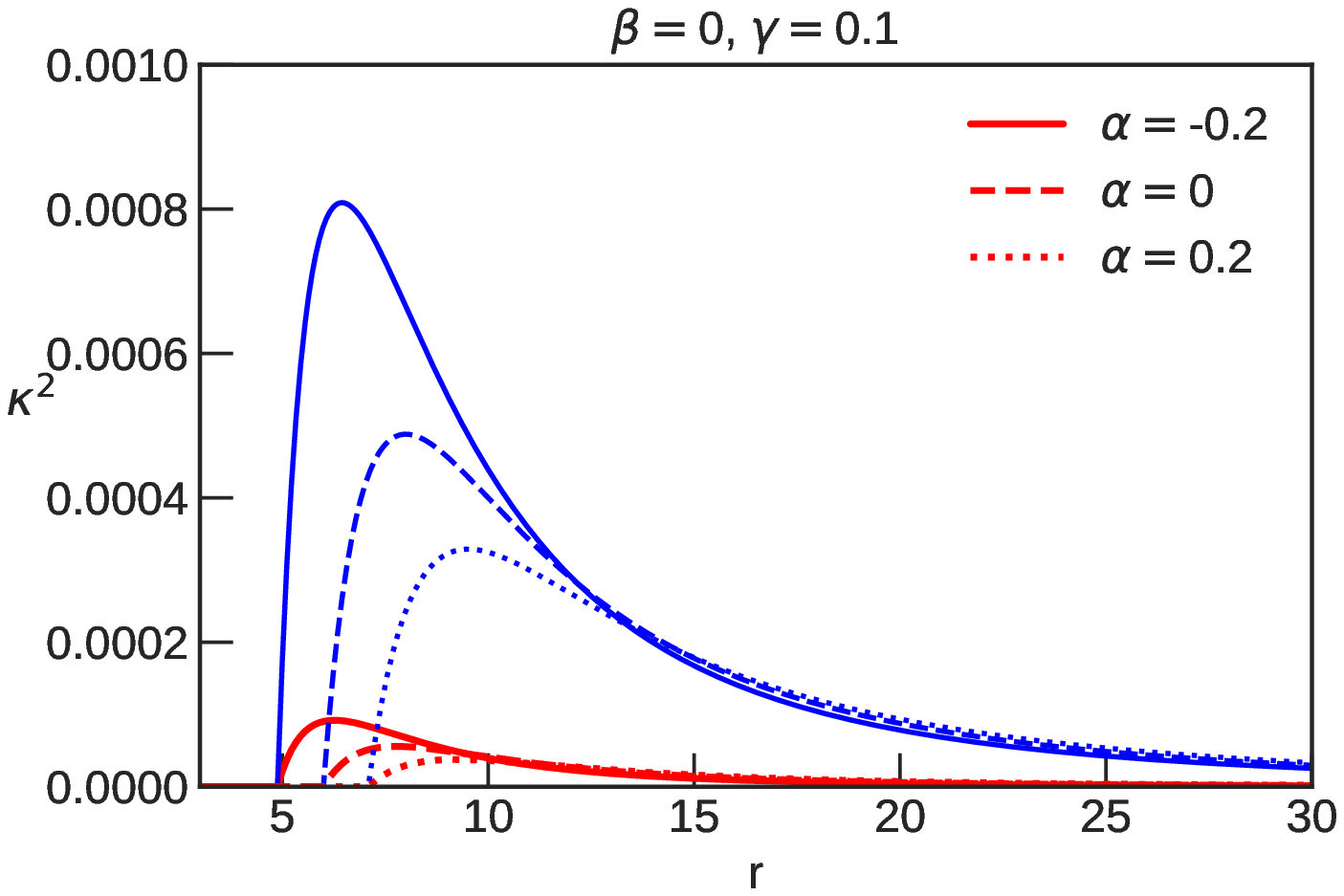}
    \includegraphics[width=0.37\hsize]{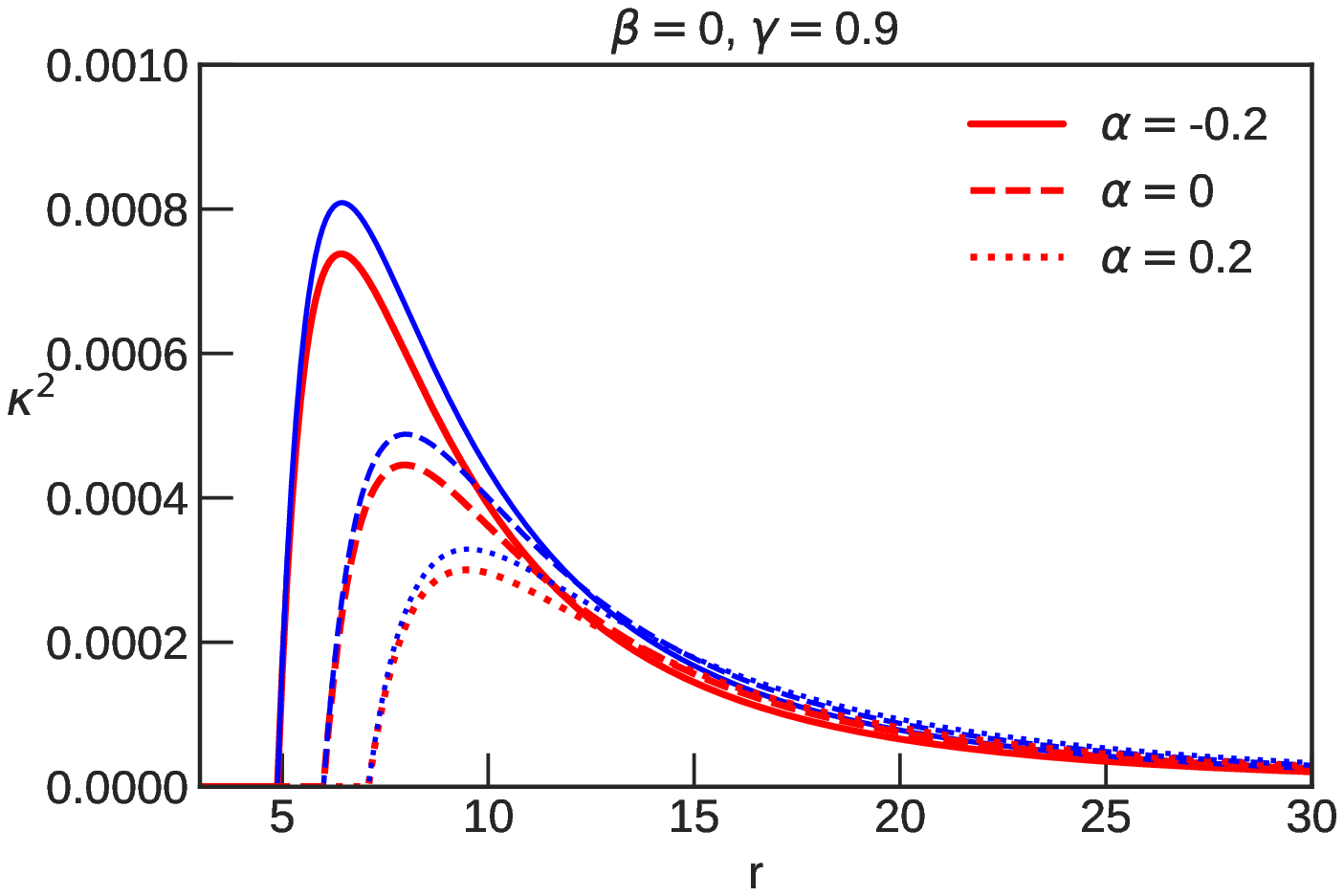}
     \end{tabular}  
    \caption{The plots of $\kappa^2$ for the tori are presented in red to compare with the blue one $\kappa^2$ for the test particle and with Keplerian angular momentum. In addition, two different parameters of $\gamma$ related to angular momentum are considered. For higher $\gamma$ we have radial epicyclic frequency approaching the Keplerian one which is compatible with its definition.}
    \label{fig:QianVSKep2}
\end{figure*}

To investigate the radial epicyclic frequency in a perfect fluid disk, considering the method described in Subsection \ref{local}, finally the relation of $\kappa$ for the tori, in this space-time is obtained as follows

\begin{align} \label{kappadisk}
    \kappa^2=\frac{2 e^{\beta(x^2-1)+2\nu-2\lambda}\left(\frac{x+1}{x-1}\right)^\alpha \Omega V\left(-2\Omega V+(x^2-1)\partial_x \Omega\right)}{F},
   \end{align}

where
\begin{align}
V&=\underbrace{1+\alpha+(\beta-1)x-\beta x^3}_{V_1}+\underbrace{(x^2-1)\partial_x \nu}_{V_2}\\
    F &= \left(x^2-1\right)\left(-e^{2\nu}+e^{\beta(x^2-1)}(x+1)^2\left(\frac{x+1}{x-1}\right)^\alpha \Omega^2\right),\nonumber\\
     \nu&=\left(\frac{1+\alpha}{2}\right)\ln{\frac{x-1}{x+1}}+\hat{\psi}\nonumber\\
     \lambda&=\left(\frac{1+\alpha}{2}\right)\ln{\frac{x+1}{x-1}}+\left(\frac{\alpha(2+\alpha)}{2}\right)\ln{\frac{x^2-1}{x^2}}-\hat{\psi}+\hat{\gamma}.\nonumber
\end{align}
If we consider the $\kappa$ squared, this radial epicyclic frequency can be written in terms of superposition of two parts. First related to the test particle and the second related to when we consider the pressure in the disk

\begin{align}\label{eqboth}
    \kappa^2=\kappa_{t}^2+\kappa_{p}^2,
   \end{align}

where 
\begin{align}
    \kappa_{t}^2=\frac{2 e^{q(x^2-1)-2\lambda}\left(\frac{x+1}{x-1}\right)^p \Omega V_1\left(2\Omega V-(x^2-1)\partial_x \Omega\right)}{(x-1)^2},
   \end{align}

\begin{align}
    \kappa_{p}^2=&\frac{
    2 e^{2q(x^2-1)-2\lambda}\left(\frac{x+1}{x-1}\right)^{p} \Omega }{F} \left[(x+1)^2\left(\frac{x+1}{x-1}\right)^{p} V_1 \Omega^2 \right. \nonumber \\
    &\left.+e^{-q(x^2-1)+2\nu}V_2 \right] \left(-2\Omega V+(x^2-1)\partial_x \Omega\right).
   \end{align}
In Figure \ref{fig:QianVSKep1} we see $\kappa^2$ plotted for the vanishing external field $\beta=0$, and for different negative, positive values of $\alpha$, and the Schwarzschild case $\alpha=0$. The parameter $\gamma$ is the angular momentum distribution variable that we discussed so far. First of all because of vertically integration that we are interested in a neighborhood of equatorial plane $\delta$ plays no role; however, we have its contribution in the disk configuration. Figure \ref{fig:QianVSKep2} presents $\kappa^2$ for two different values of $\gamma$; in addition, we can compare the $\kappa^2$ for a fluid constructed by a non-constant angular momentum depicted in red (equation \eqref{osyqianangular}) with $\kappa^2$ for a test particle with the Keplerian angular momentum in blue. In general, larger parameter $\gamma$ has more influence as we go far from the central object which is compatible with its definition \eqref{osyqianangular}. Besides, considering Figures \ref{fig:QianVSKep1} and \ref{fig:QianVSKep2} together reveal the effect of different angular momentum on the radial epicyclic frequency of a test particle, which is consistent with what we saw from Figures \ref{fig:MapQianpq} and \ref{fig:MapConstpq}.

\begin{figure}
\centering
         \includegraphics[width=0.66\hsize]{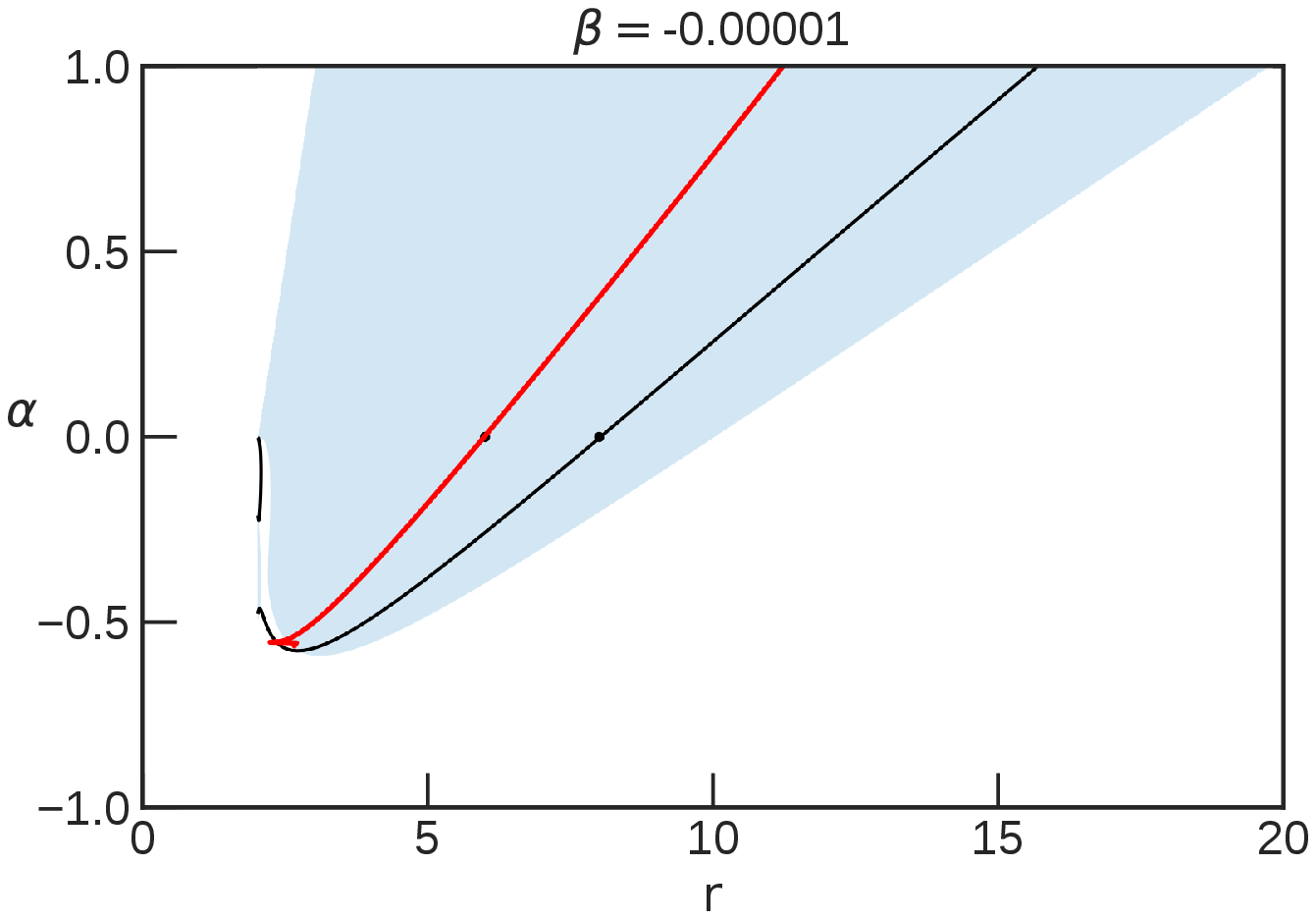}
         \includegraphics[width=0.72\hsize]{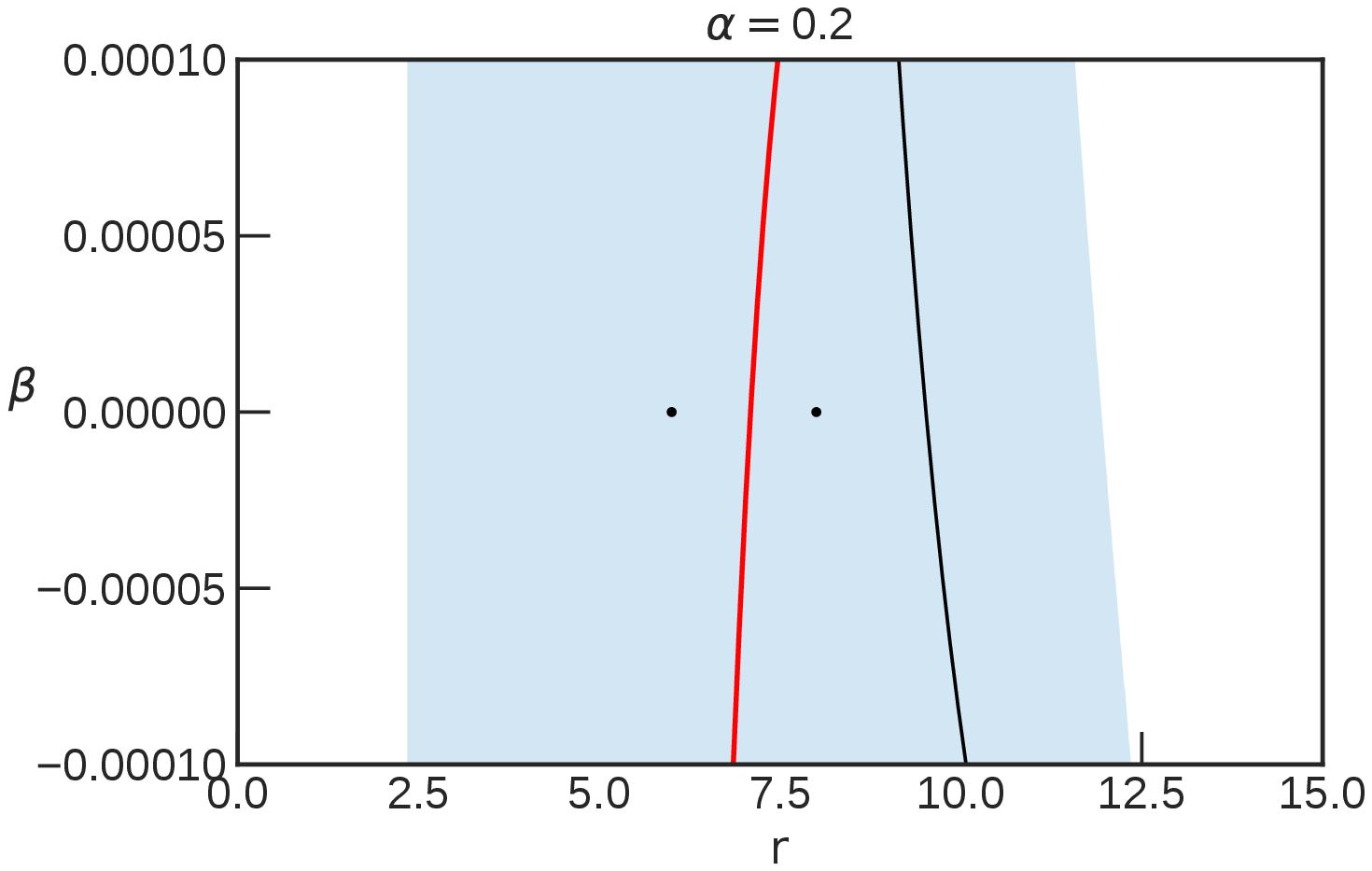}
   \caption{The red line presents the place of ISCO with respect to $\alpha$ and $\beta$. The black line depicts the value of the maxima of $\kappa$. The blue region shows where $\kappa$ can exhibit maxima. The two dots give, from left to right, ISCO and the maximum of $\kappa$ for the Schwarzschild case. 
   }
    \label{fig:condp}
\end{figure}



\begin{figure*}
    \centering
    \begin{tabular}{ccc}
      \includegraphics[width=0.33\hsize]{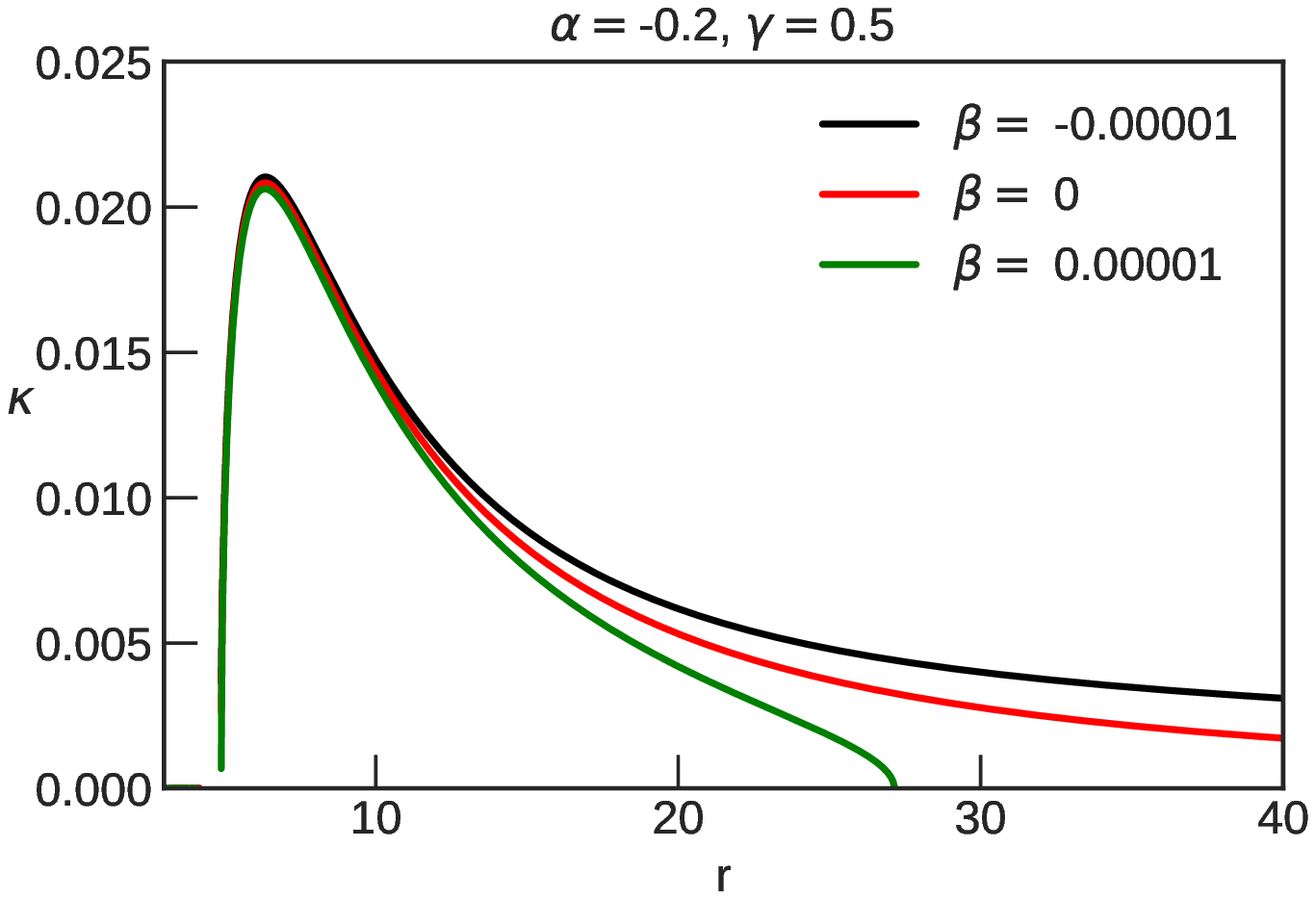} &
     \includegraphics[width=0.33\hsize]{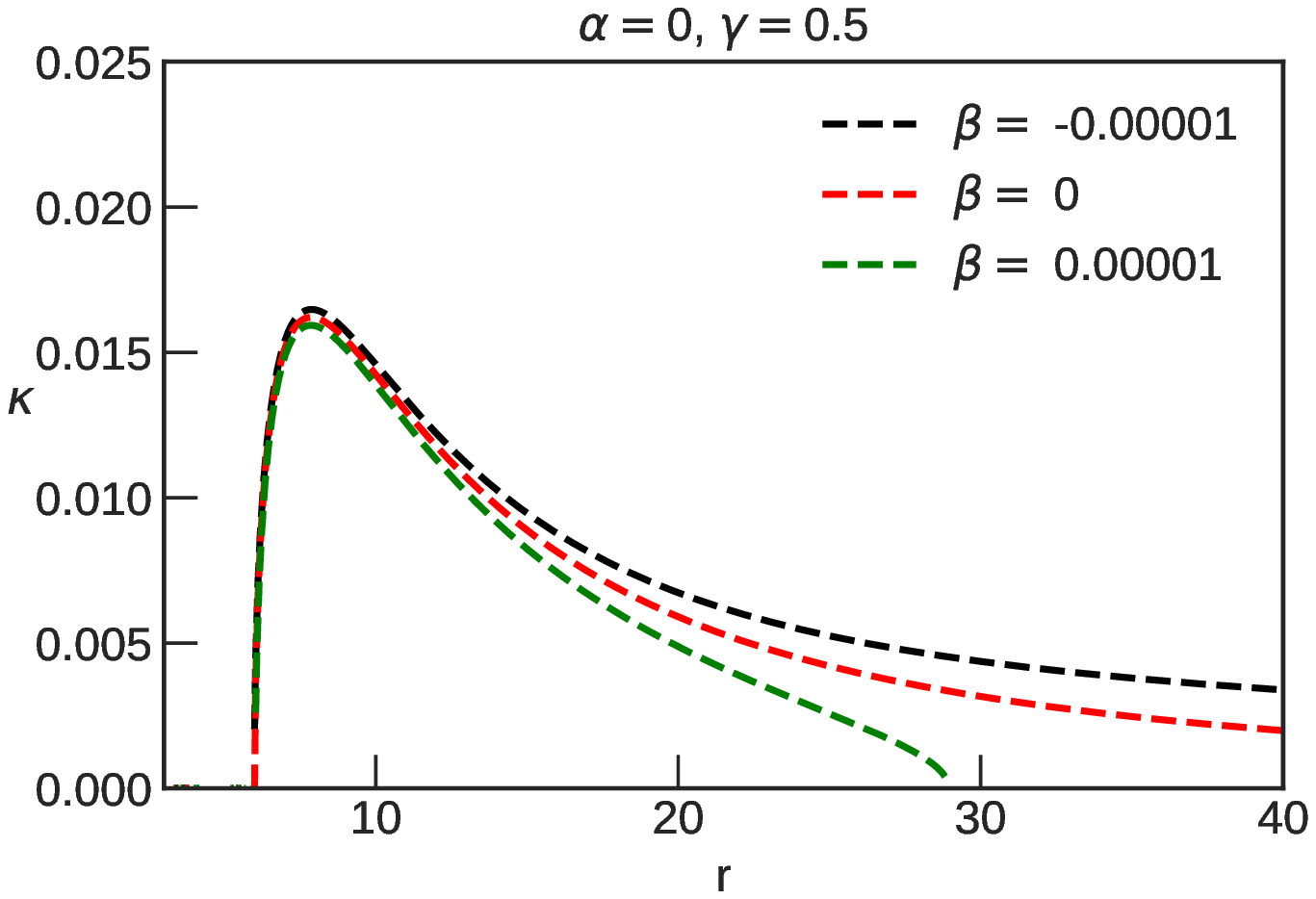} & \includegraphics[width=0.33\hsize]{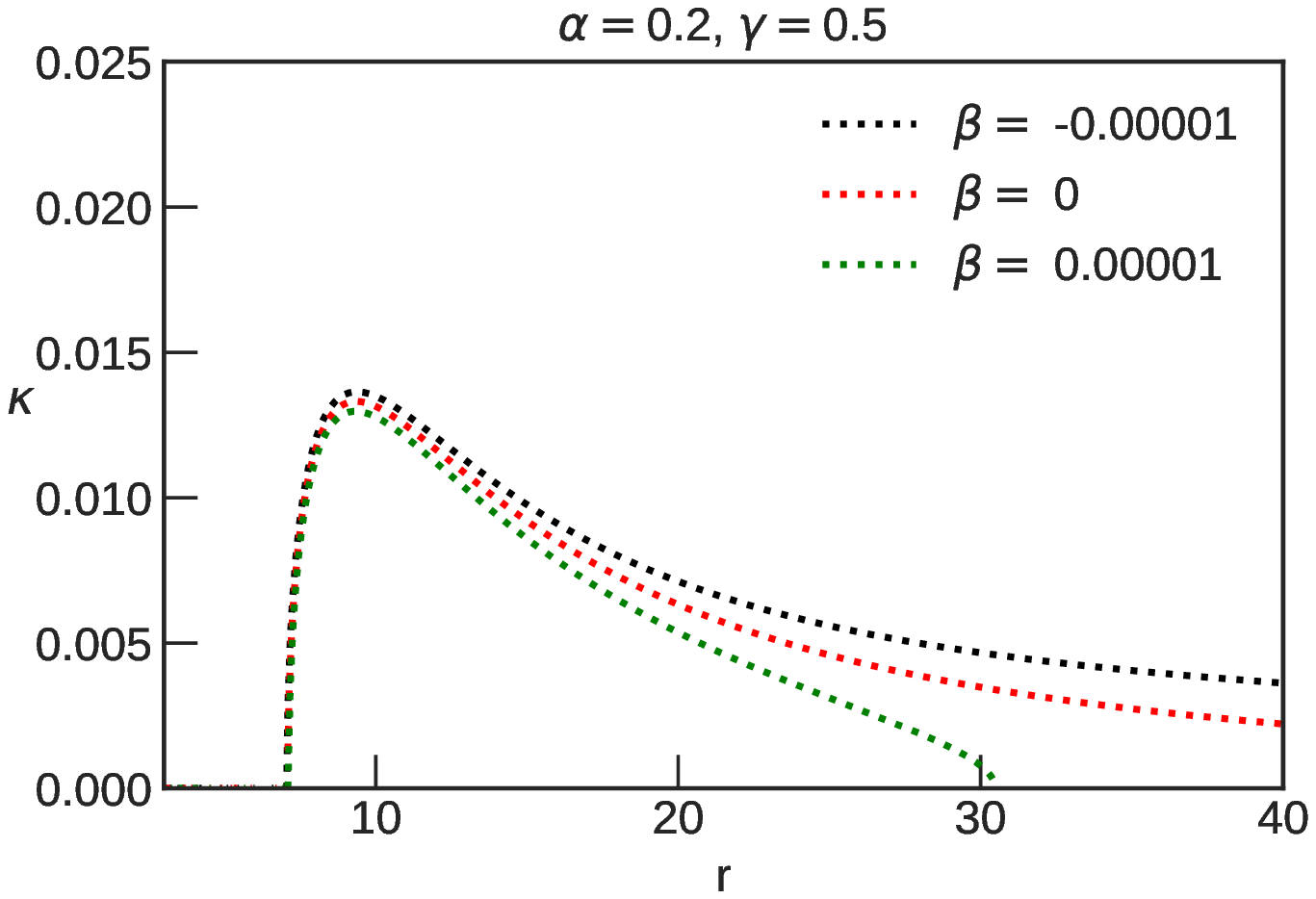}\\
     \end{tabular}  
    \caption{The plots of $\kappa$ for tori with respect to $r$ for various combination of $\beta$, $\alpha$ and $\gamma$.} 
    \label{fig:QianVSKep2fluid}
\end{figure*}

The important key here is in general theory of relativity $\kappa$ does not increase as we go closer and closer to the central object, but instead have a maximum at a few gravitational radii before the innermost stable circular orbit (ISCO) where consider as a very good estimation of the inner edge of the disc, and it is zero at the ISCO. Therefore, oscillations in this part of the disk may trap and lead to the periodic flux variations which we can observe \cite{1980PASJ...32..377K}. However, this is not the case in the Newtonian gravity. For instance, Figure \ref{fig:condp} for different range of quadrupole parameters shows the maximum of $\kappa$ appears before the place of ISCO in each case. It is worth mentioning that the place of ISCO is dependent on quadrupole moments and this is closer to the central source for negative ones \cite{2020arXiv201015723F}. In this Figure, the red line depicted the place of ISCO and the black line the place of $\kappa$ maxima for a test particle.

One can see the influence of different sign of parameter $\beta$, as well as $\alpha$ on the radial epicyclic frequency $\kappa$ for tori in Figure \ref{fig:QianVSKep2fluid}. We see that for both $\alpha<0$ and $\beta$ the maximum is higher than for their zero and positive values, also it is more closer to the central object. Furthermore, there is an interesting fact about positive quadrupole $\beta$. When $\beta$ is positive the radial epicylic frequency terminates much faster regardless of sign of $\alpha$. As we discuss it in Newtonian picture, this is the manifestation of the external field when the ring like part is greater.

\subsection{QPO models and observation}

 \begin{figure*}
    \centering
    \begin{tabular}{ccc}
    \includegraphics[width=0.33\hsize]{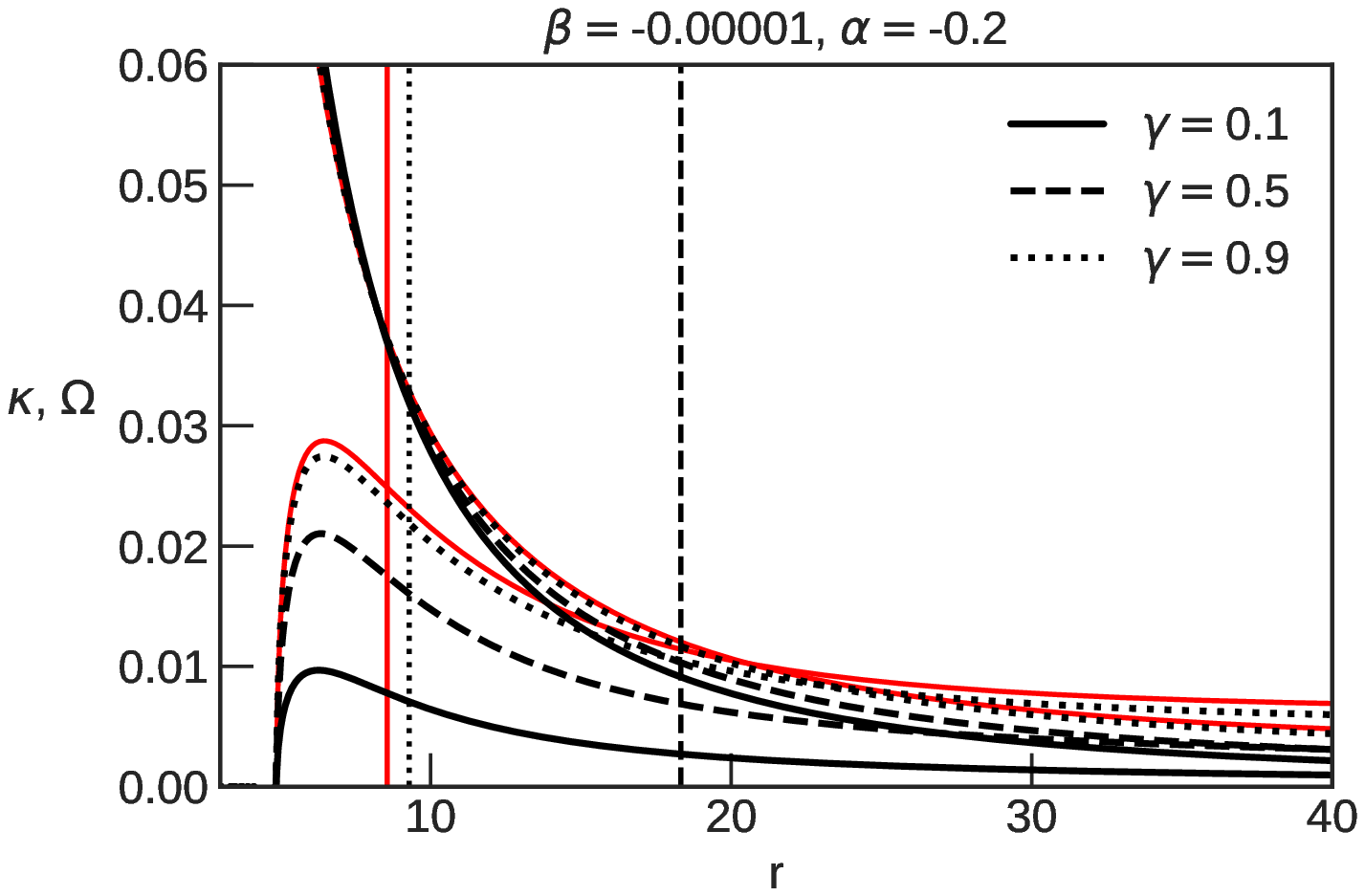} &
     \includegraphics[width=0.33\hsize]{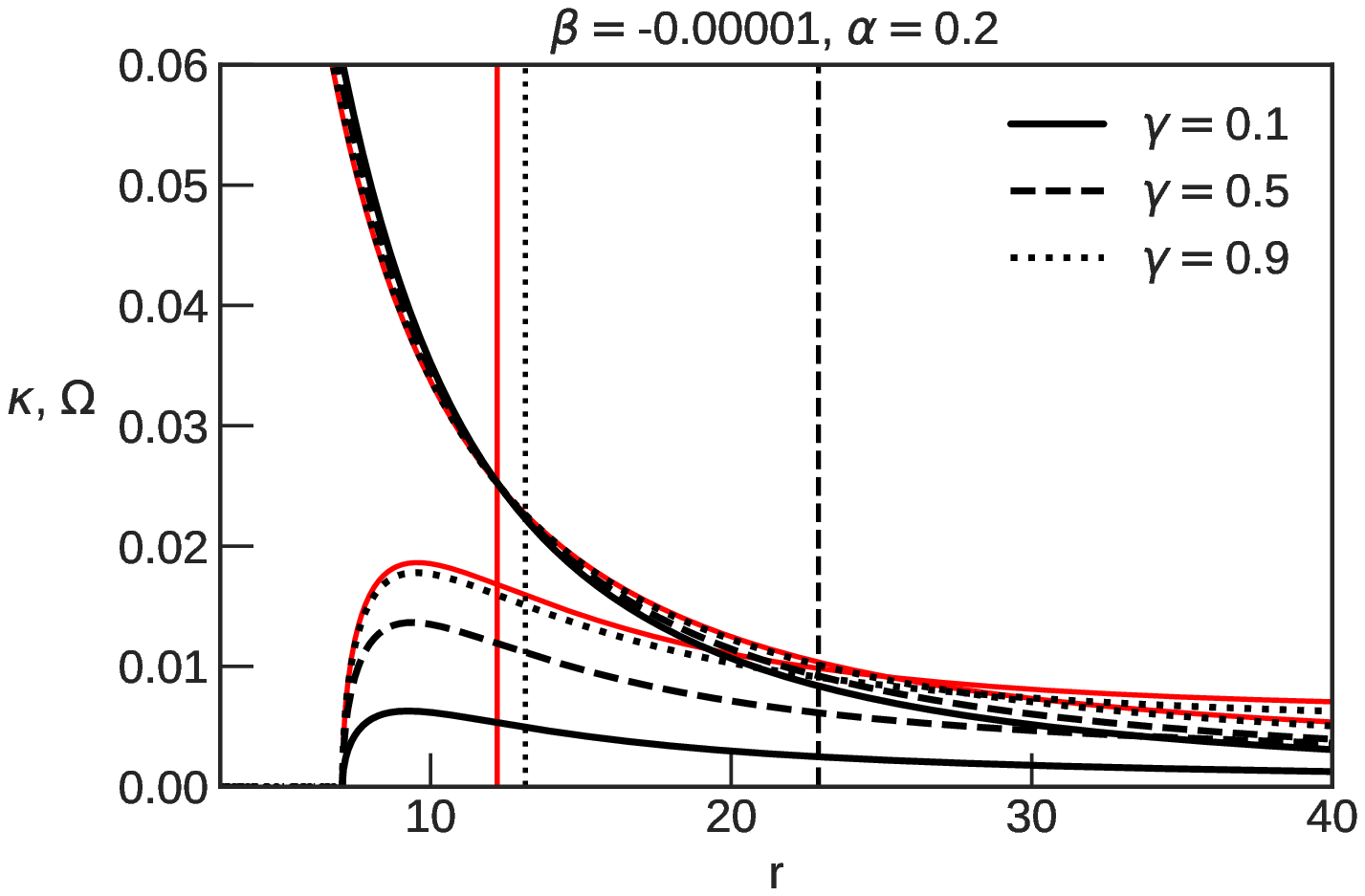} \\
     \includegraphics[width=0.33\hsize]{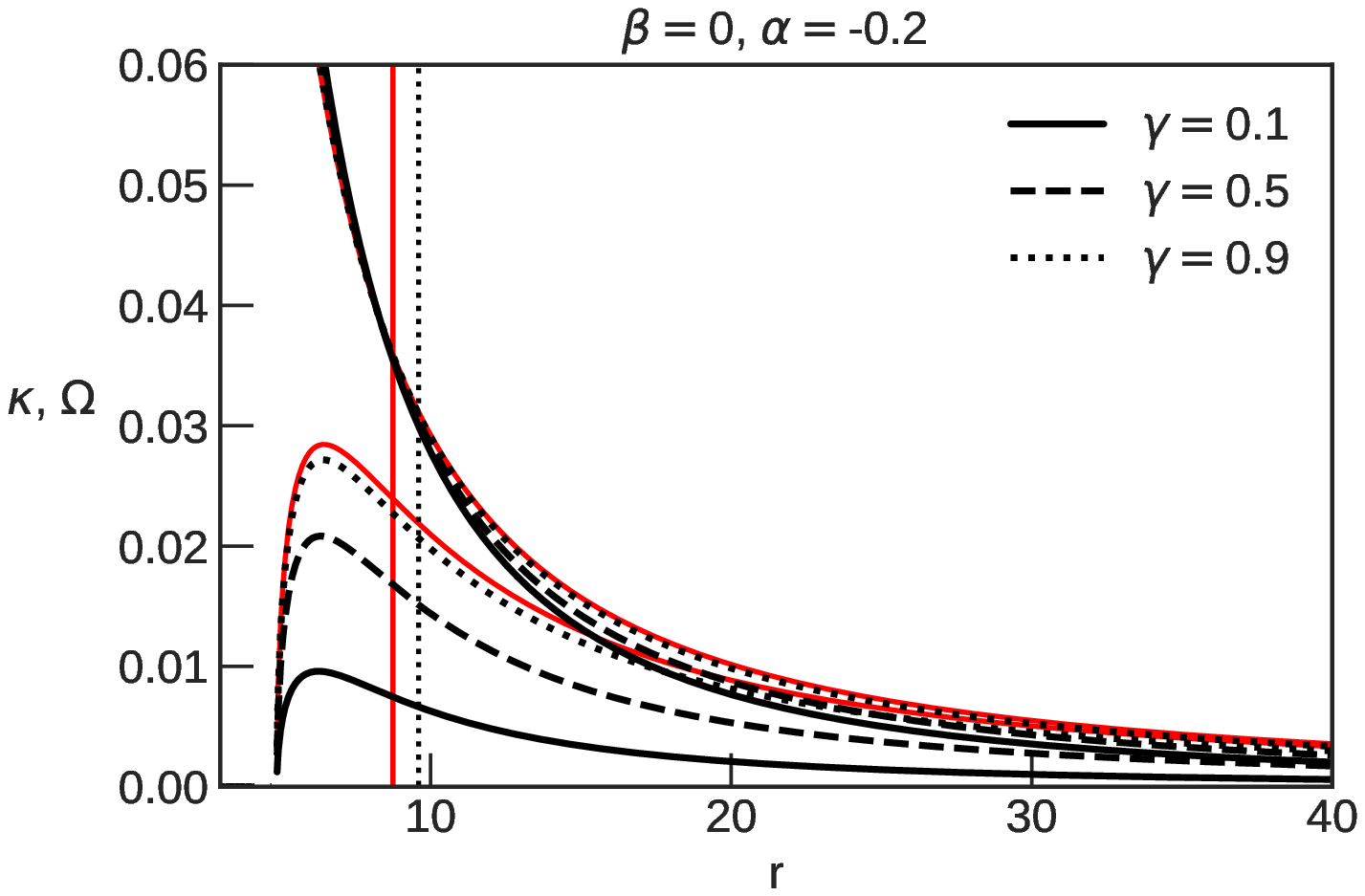} &
       \includegraphics[width=0.33\hsize]{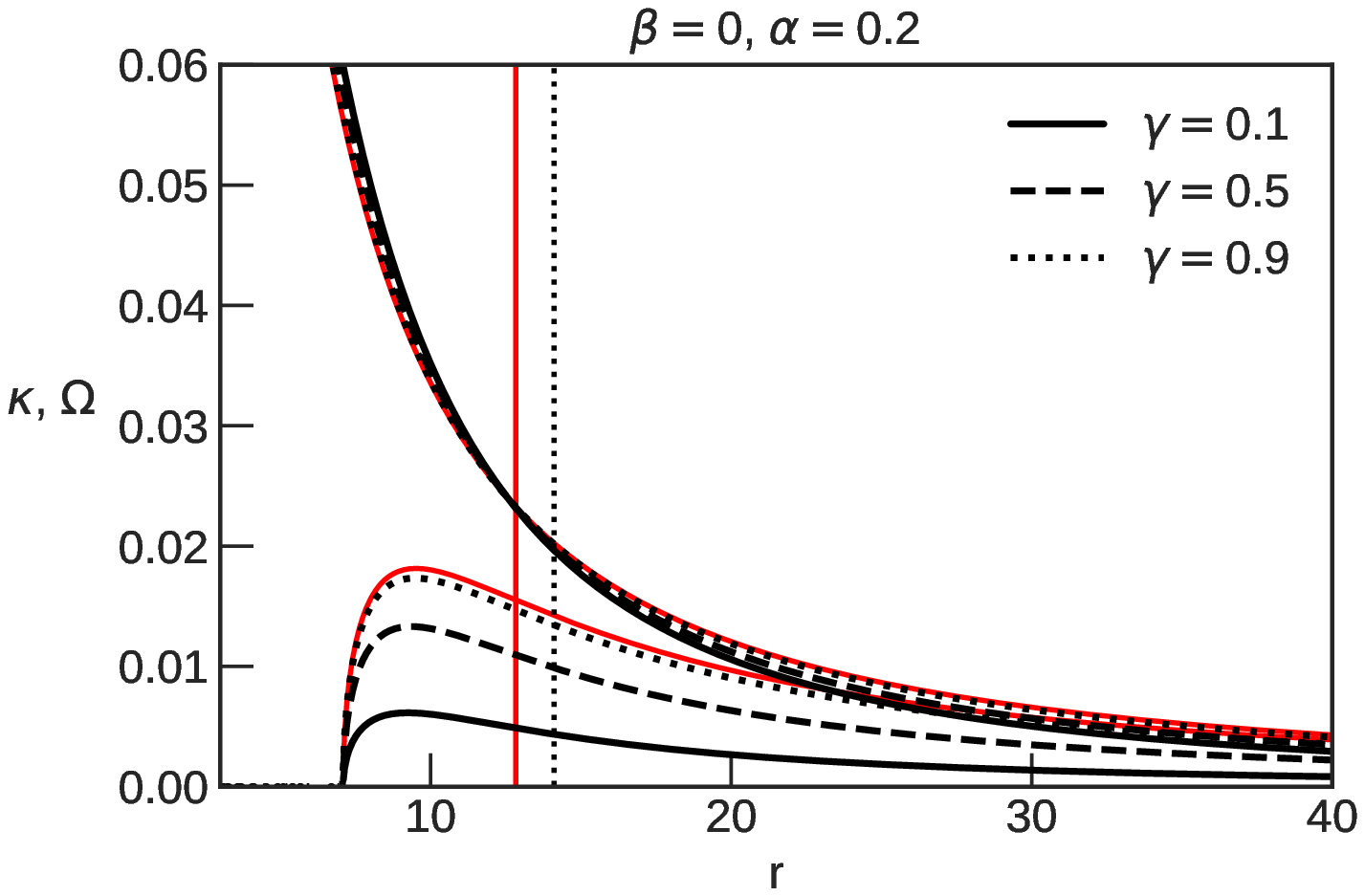} \\
     \includegraphics[width=0.33\hsize]{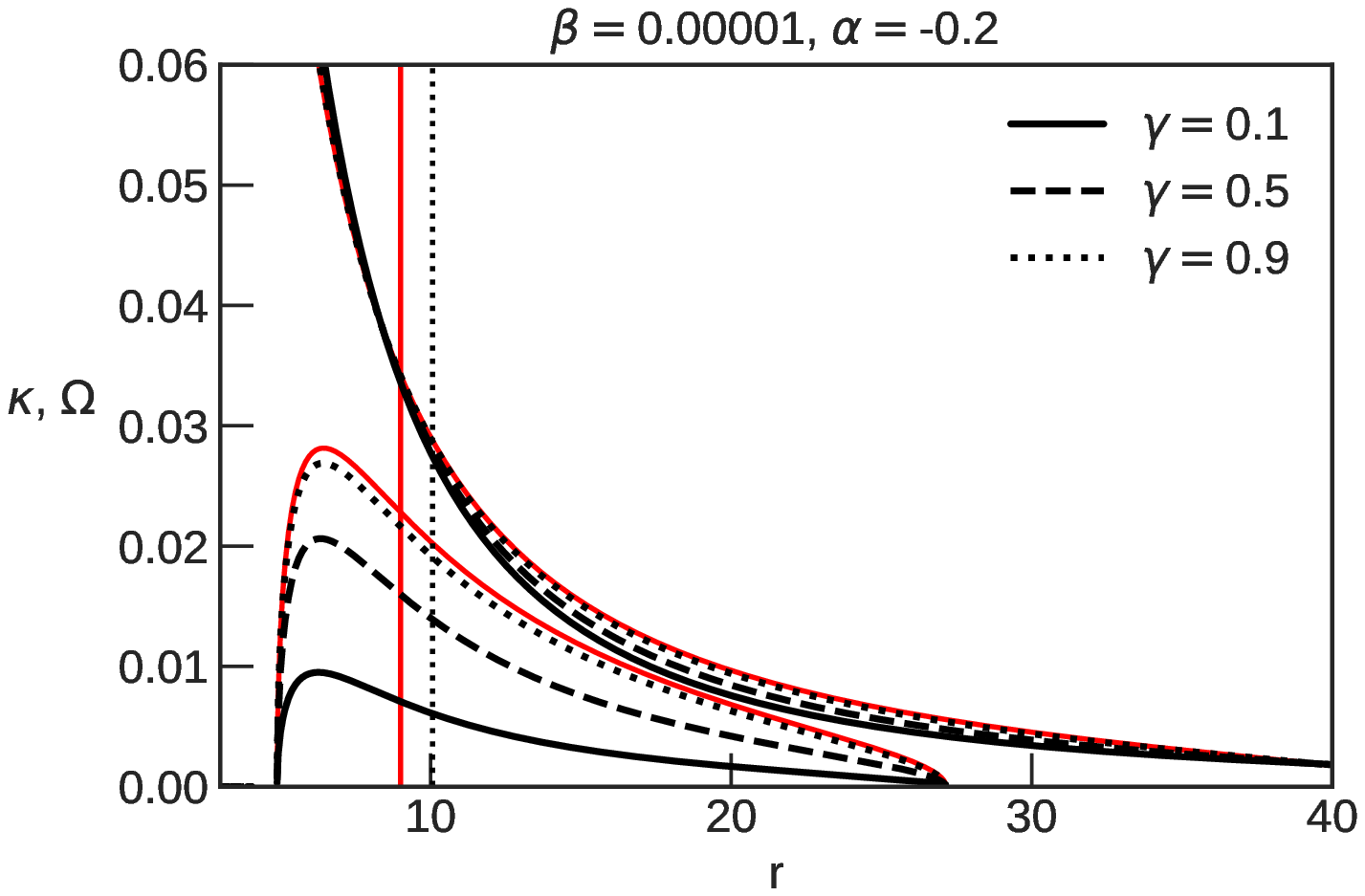} &
     \includegraphics[width=0.33\hsize]{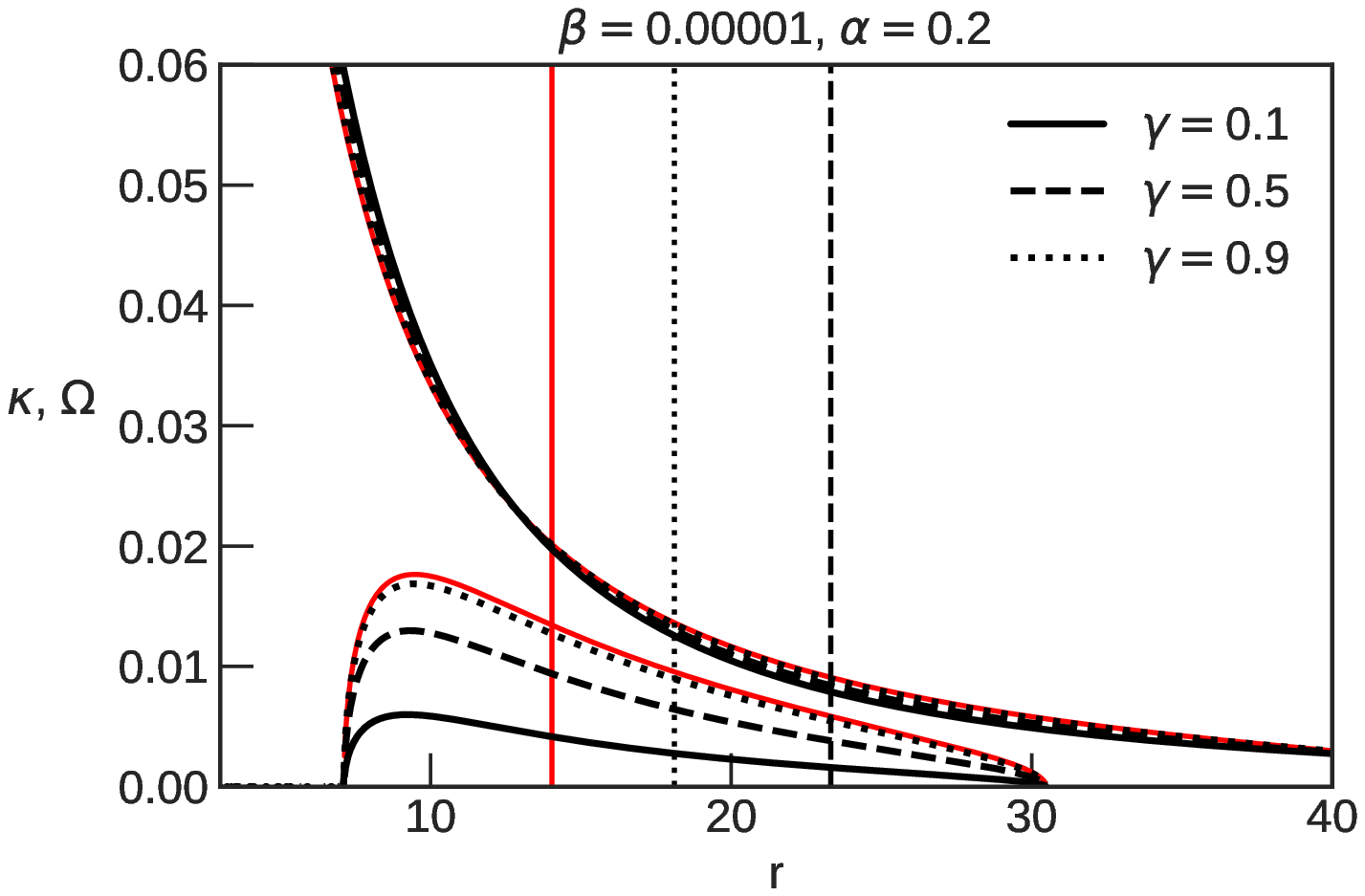} 
     \end{tabular}  
    \caption{$\kappa$ and $\Omega$ (monotonic curves) with respect to $r$ for tori and different values of $\alpha$, $\beta$ and $\gamma$. The red lines are respectively $\kappa$ and $\Omega$ for the test particle case.The vertical line corresponds to the radius of the $3:2$ ratio between $\Omega : \kappa$ for KP model.} 
    \label{fig:QianVSKep210}
\end{figure*}



\begin{figure*}
    \centering
    \begin{tabular}{cc}
    \includegraphics[width=0.33\hsize]{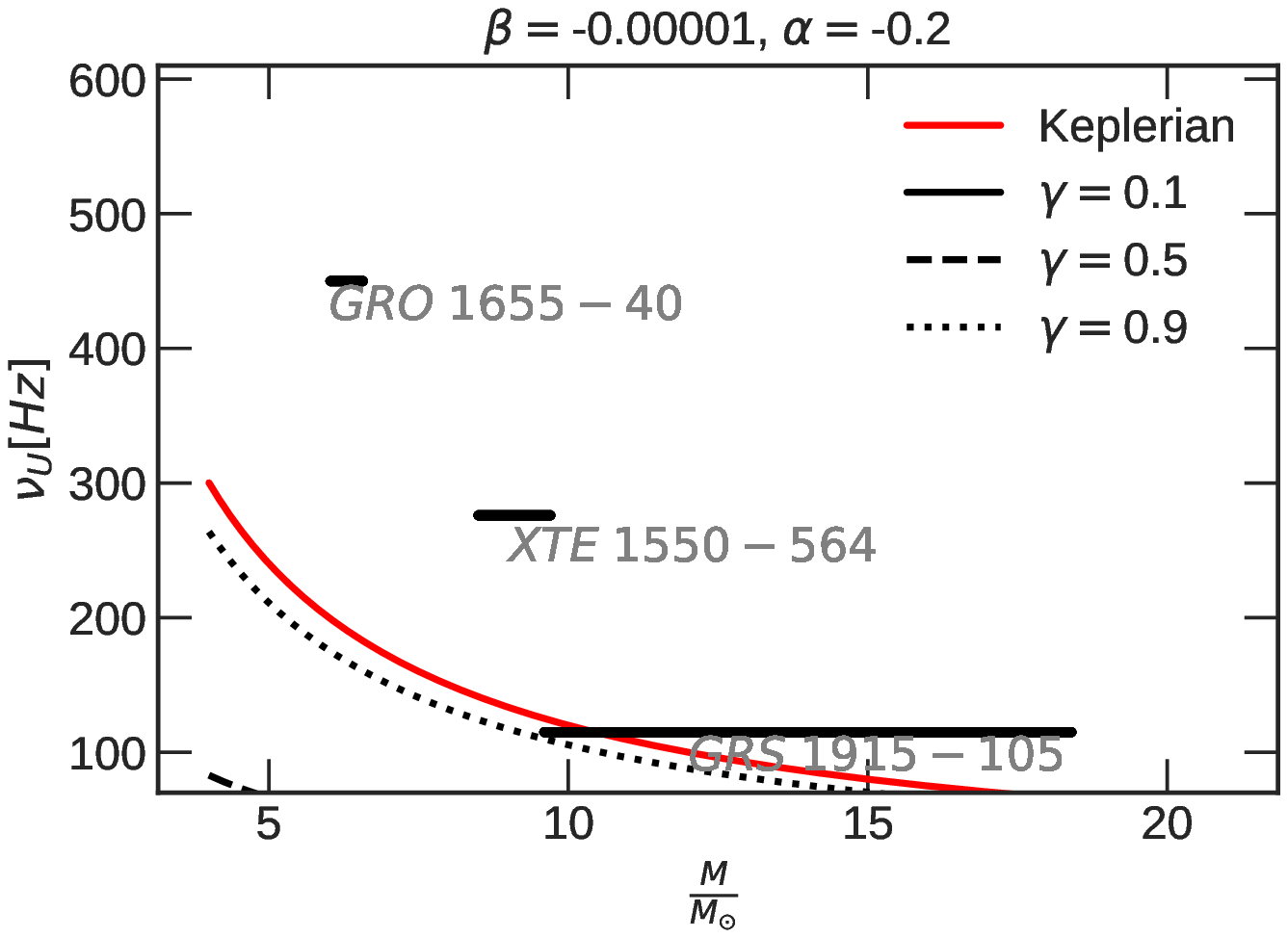} &
     \includegraphics[width=0.33\hsize]{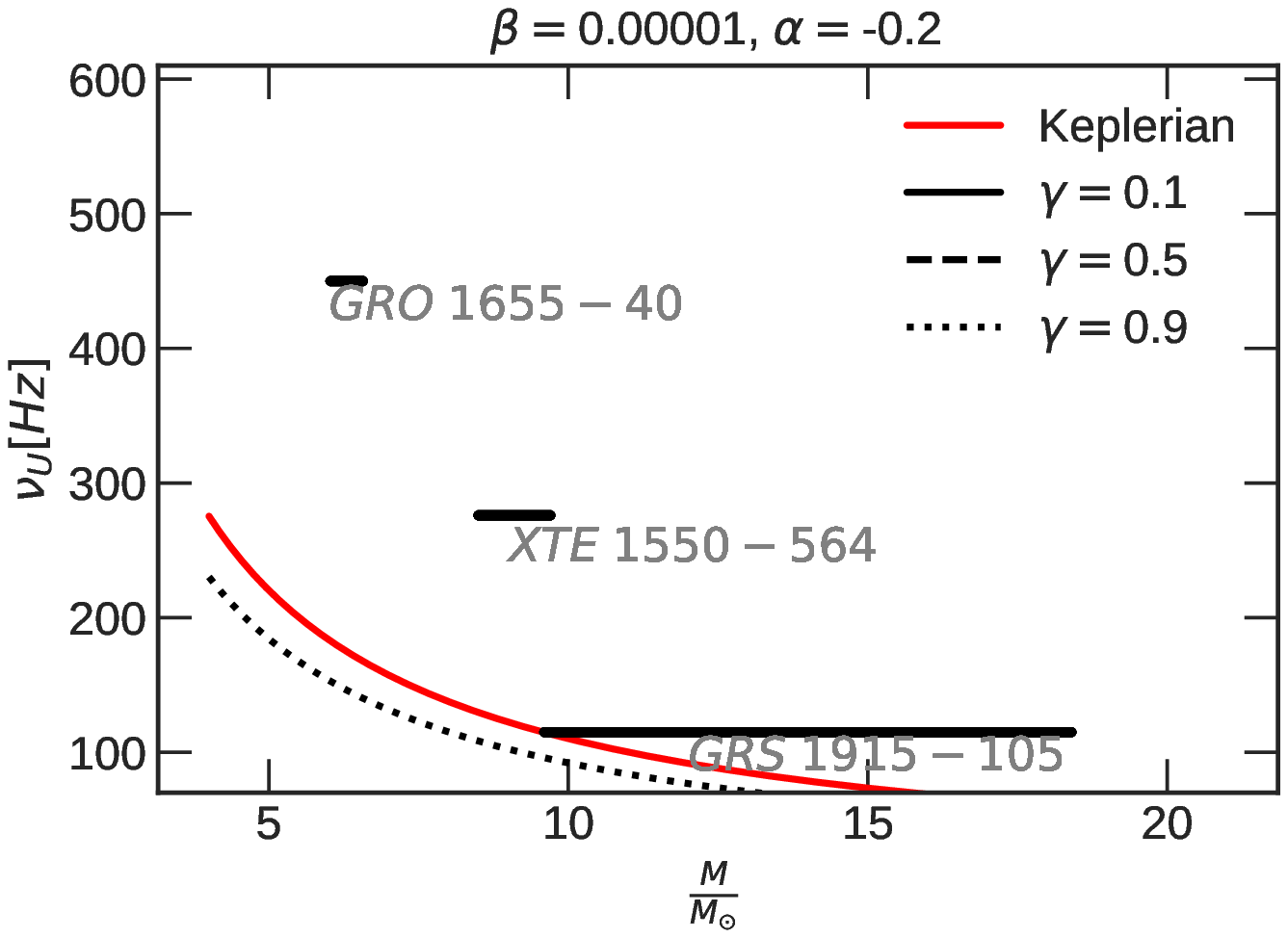} \\
     \includegraphics[width=0.33\hsize]{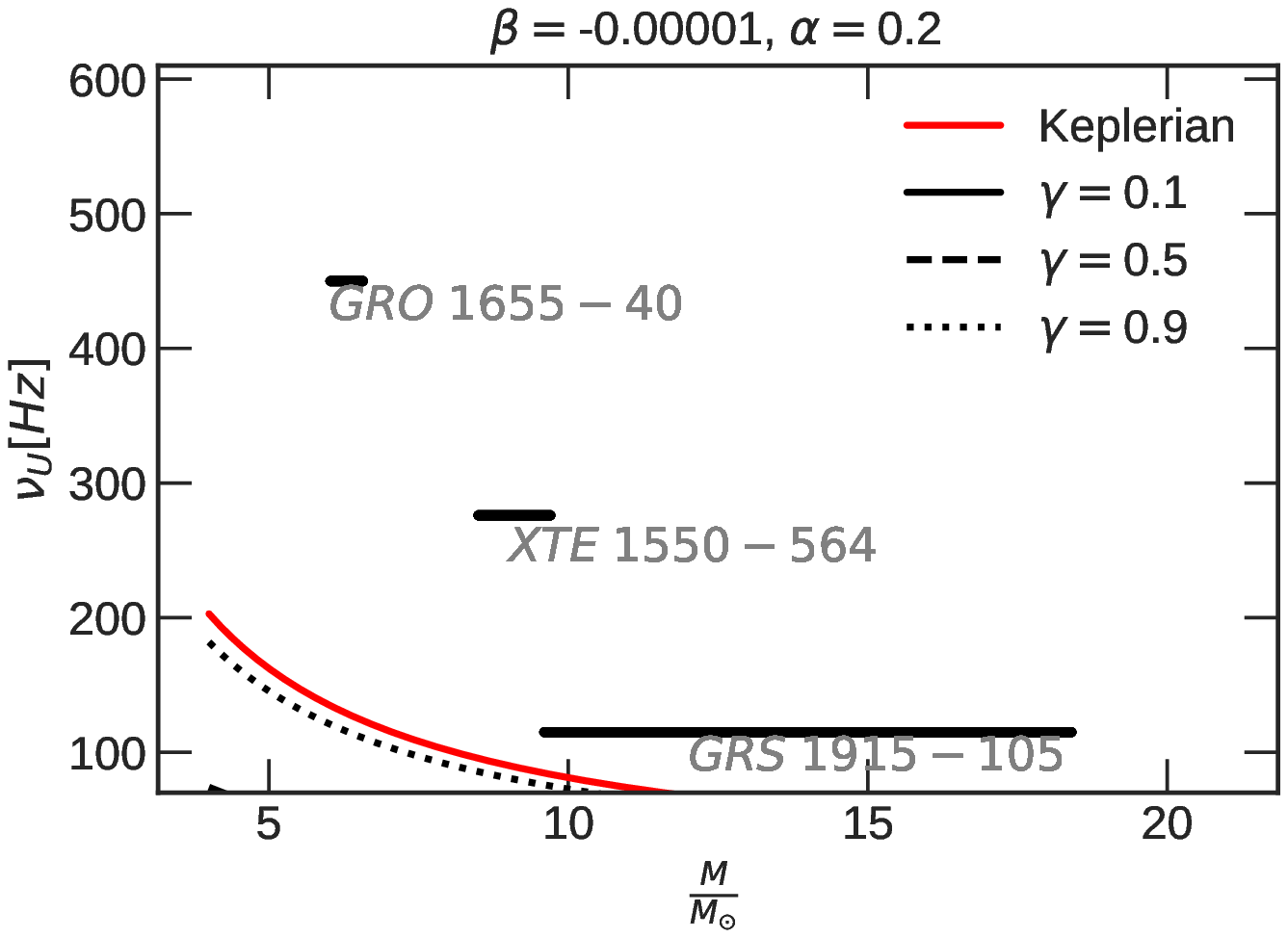} &
     \includegraphics[width=0.33\hsize]{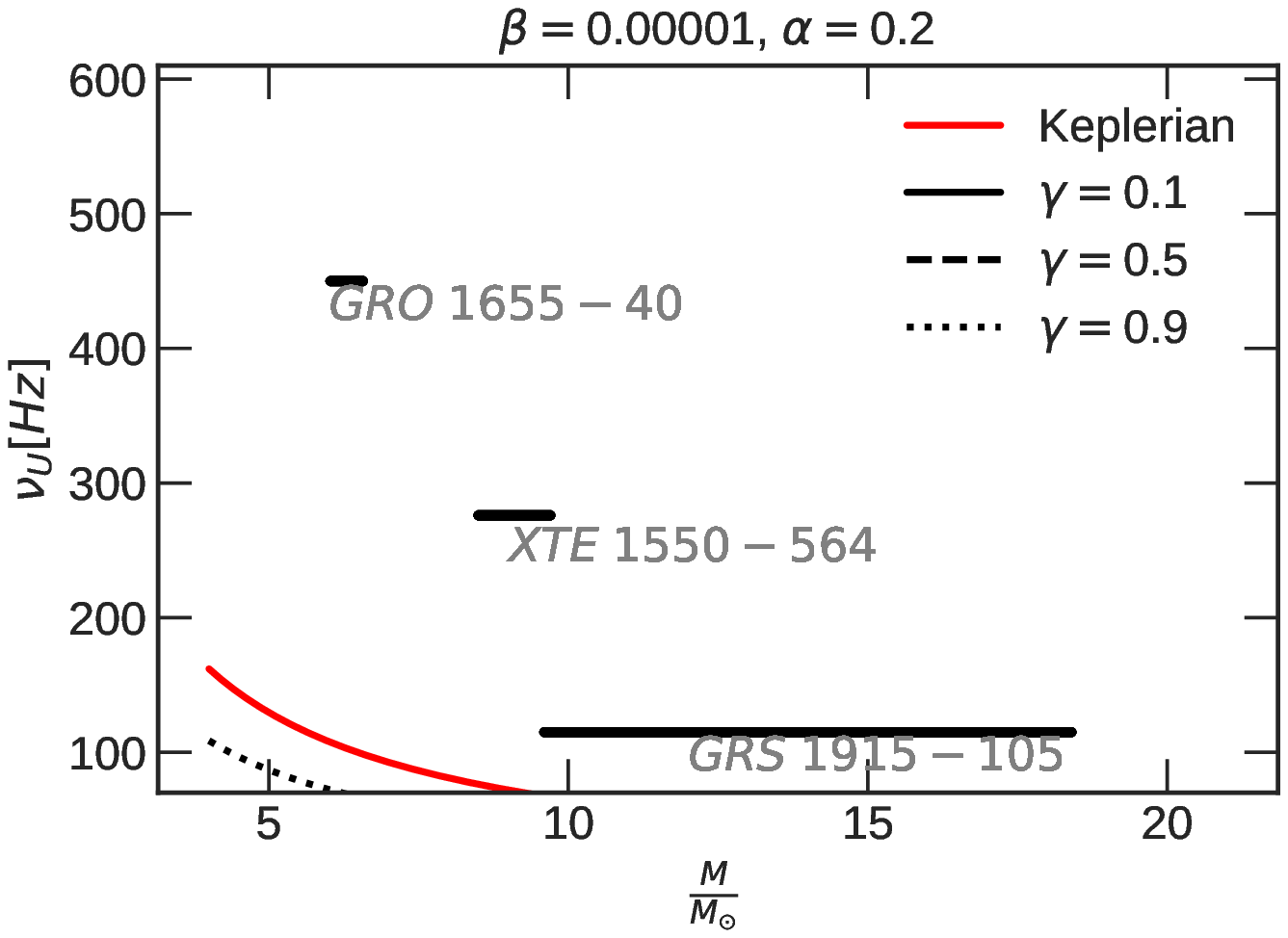} 
     \end{tabular}  
    \caption{The upper oscillation frequency $\nu_U$ at the resonance radius $3:2$ is presented for various combinations of the studied parameters for the KP model. For the test-particle is depicted in red. The black lines present the fluid for various angular momentum. If the linestyle does not appear, it means that the chosen ratio $3:2$ is not possible.} 
    \label{fig:KPModel}
\end{figure*}

\begin{figure*}
    \centering
    \begin{tabular}{cc}
    \includegraphics[width=0.33\hsize]{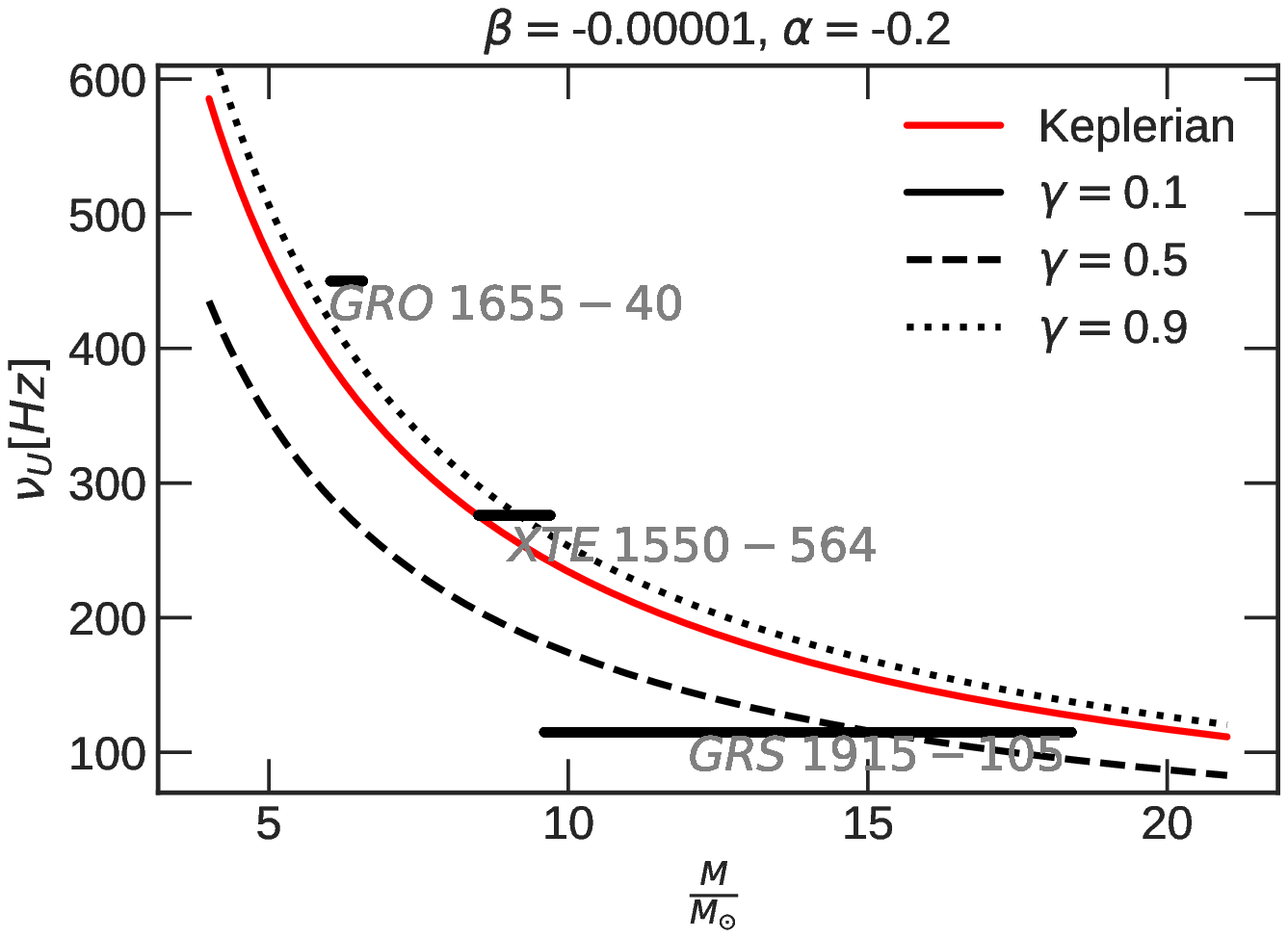} &
     \includegraphics[width=0.33\hsize]{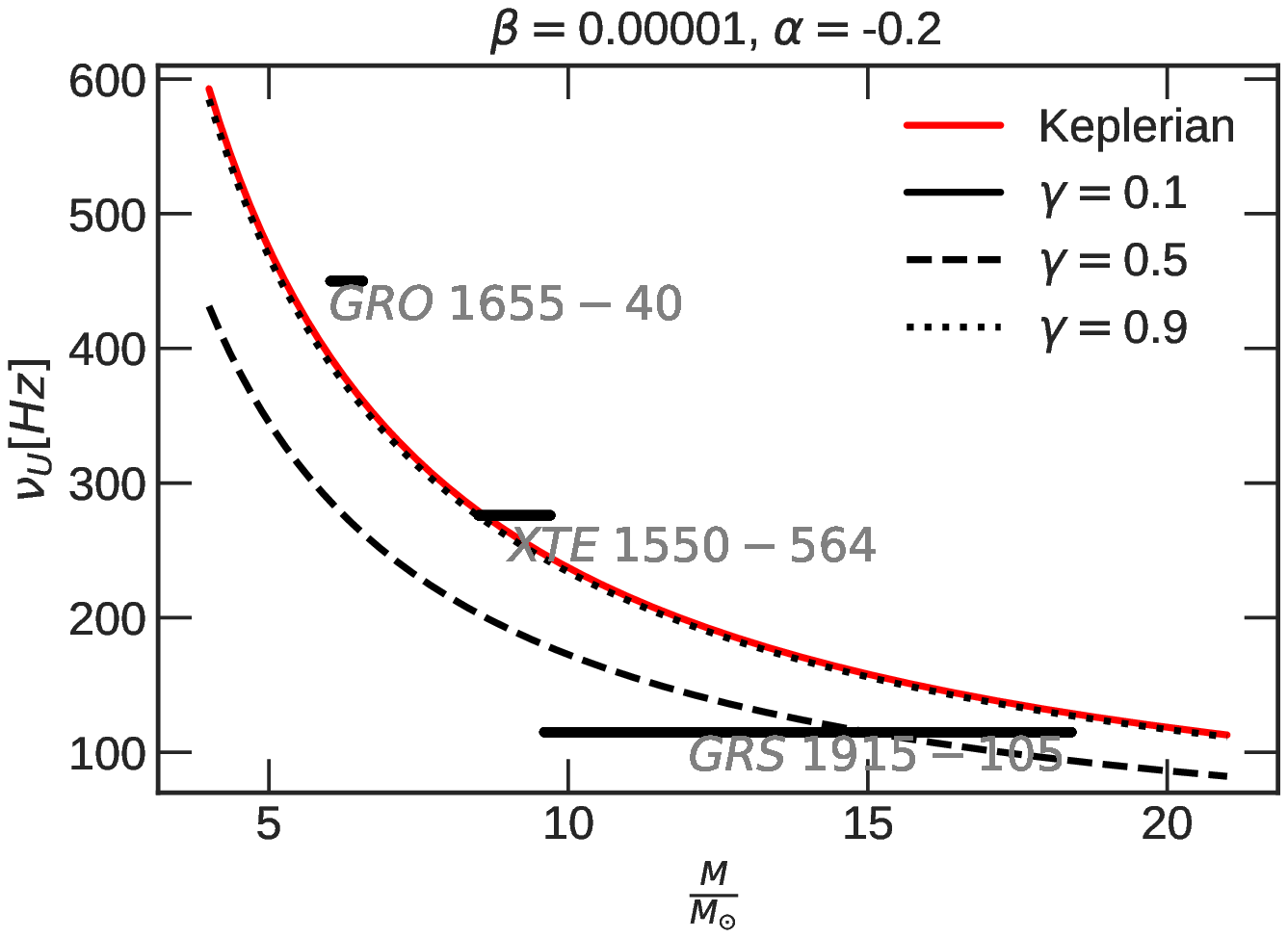} \\
     \includegraphics[width=0.33\hsize]{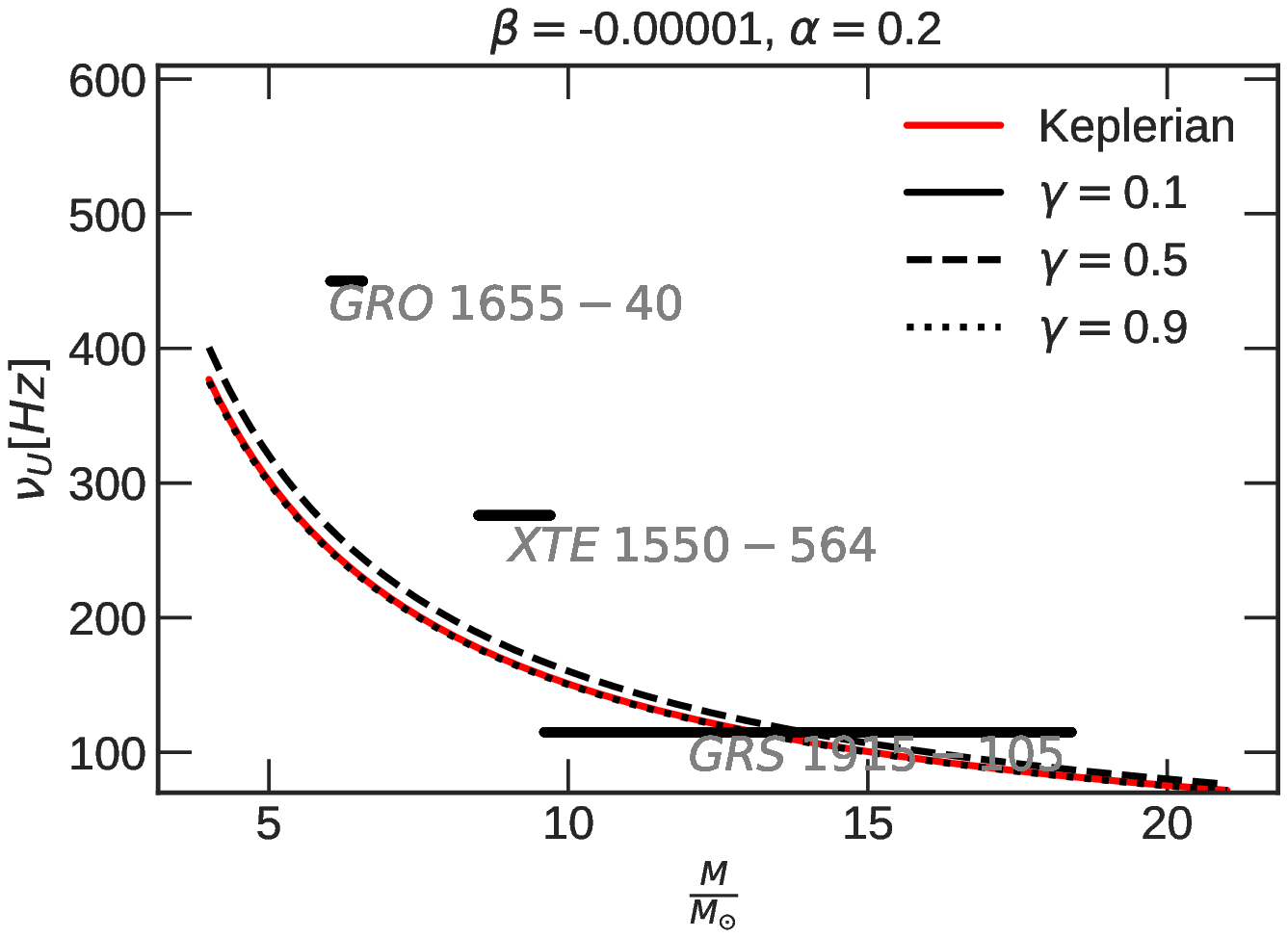} &
     \includegraphics[width=0.33\hsize]{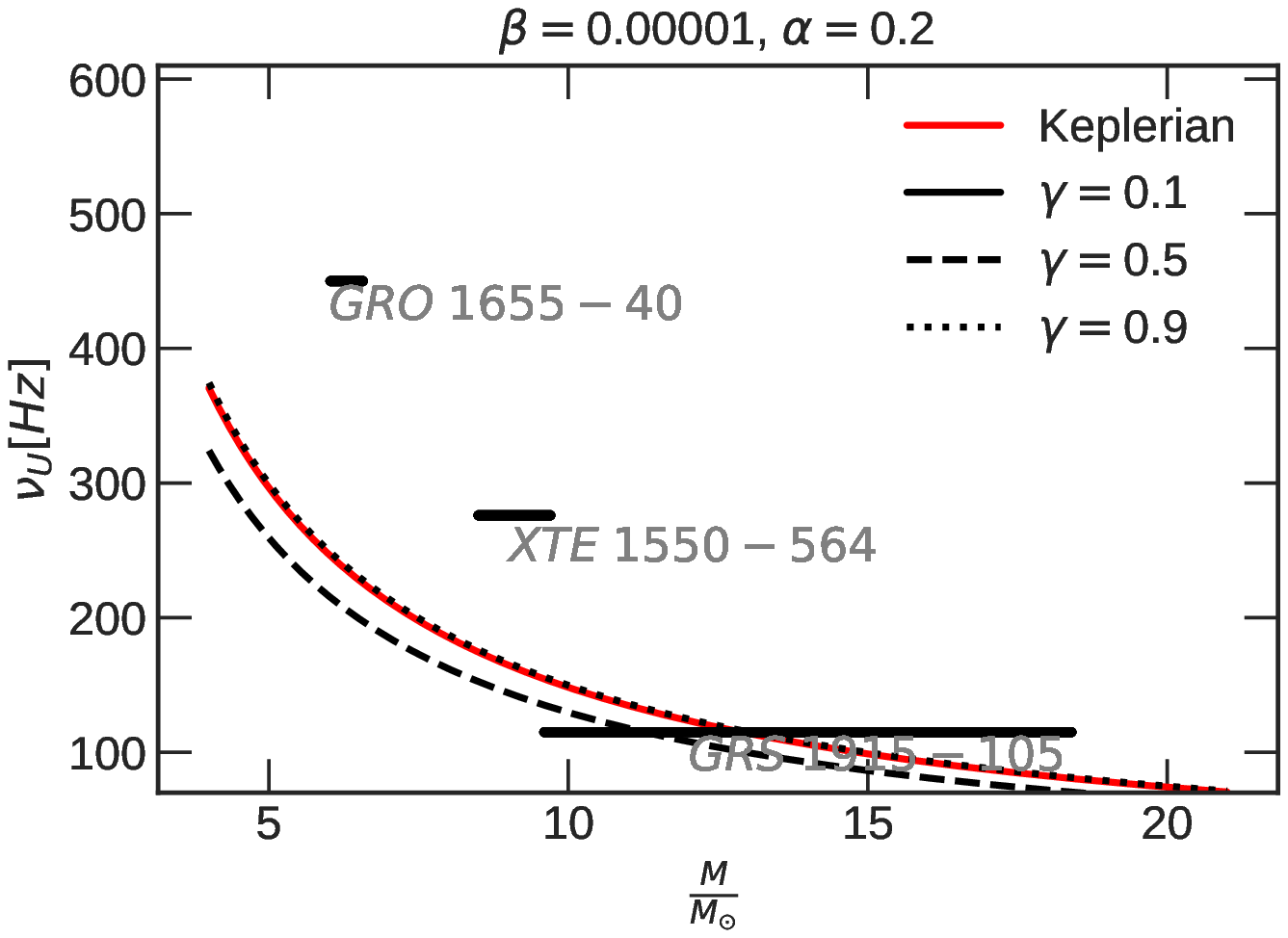} 
     \end{tabular}  
    \caption{The upper oscillation frequency $\nu_U$ at the resonance radius $3:2$ is presented for various combinations of the studied parameters for the RP model. For the test-particle is depicted in red. The black lines present the fluid for various angular momentum. If the linestyle does not appear, it means that the chosen ratio $3:2$ is not possible.}  
    \label{fig:RPmodel}
\end{figure*}

\begin{figure*}
    \centering
    \begin{tabular}{cc}
    \includegraphics[width=0.33\hsize]{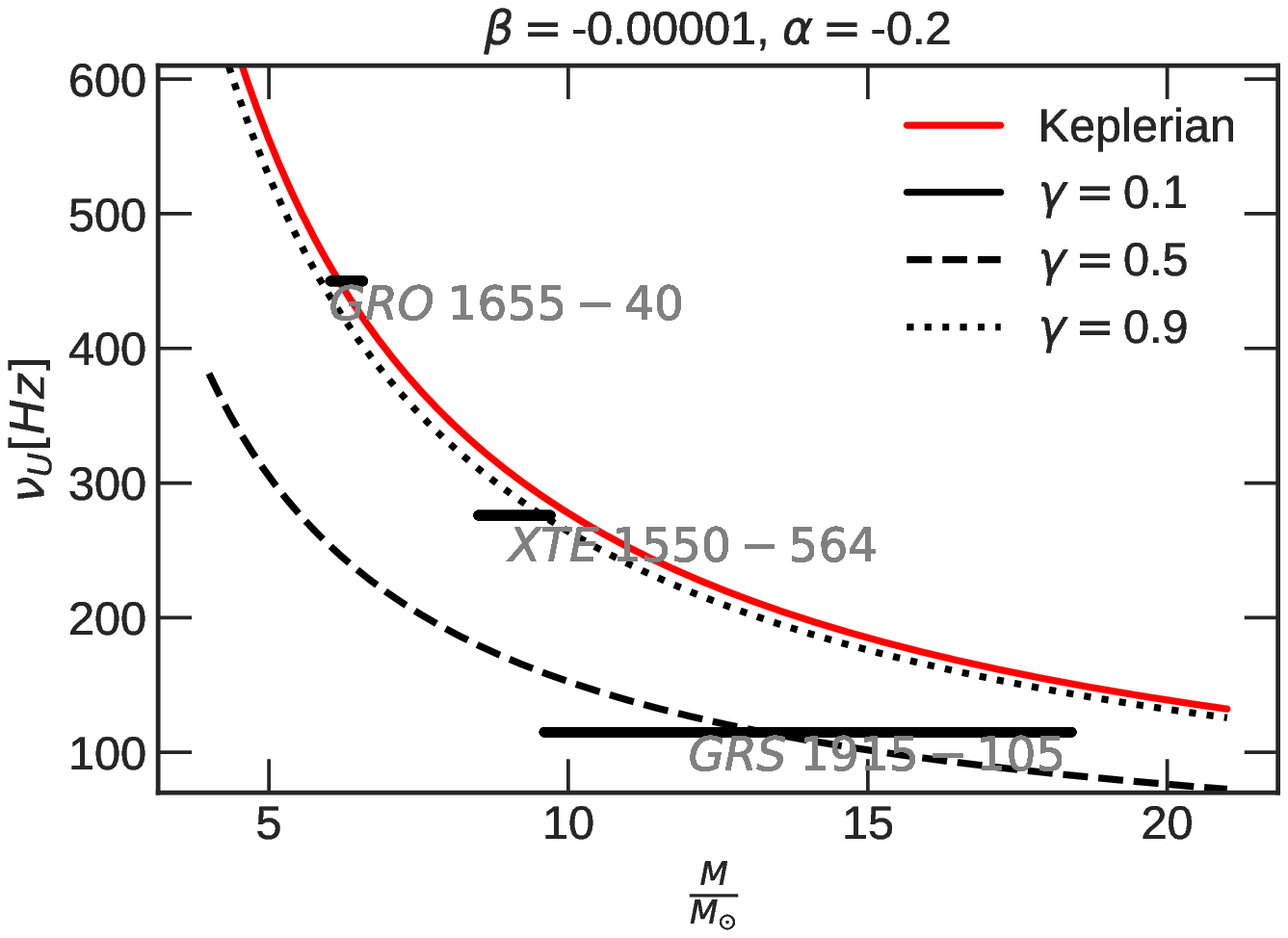} &
     \includegraphics[width=0.33\hsize]{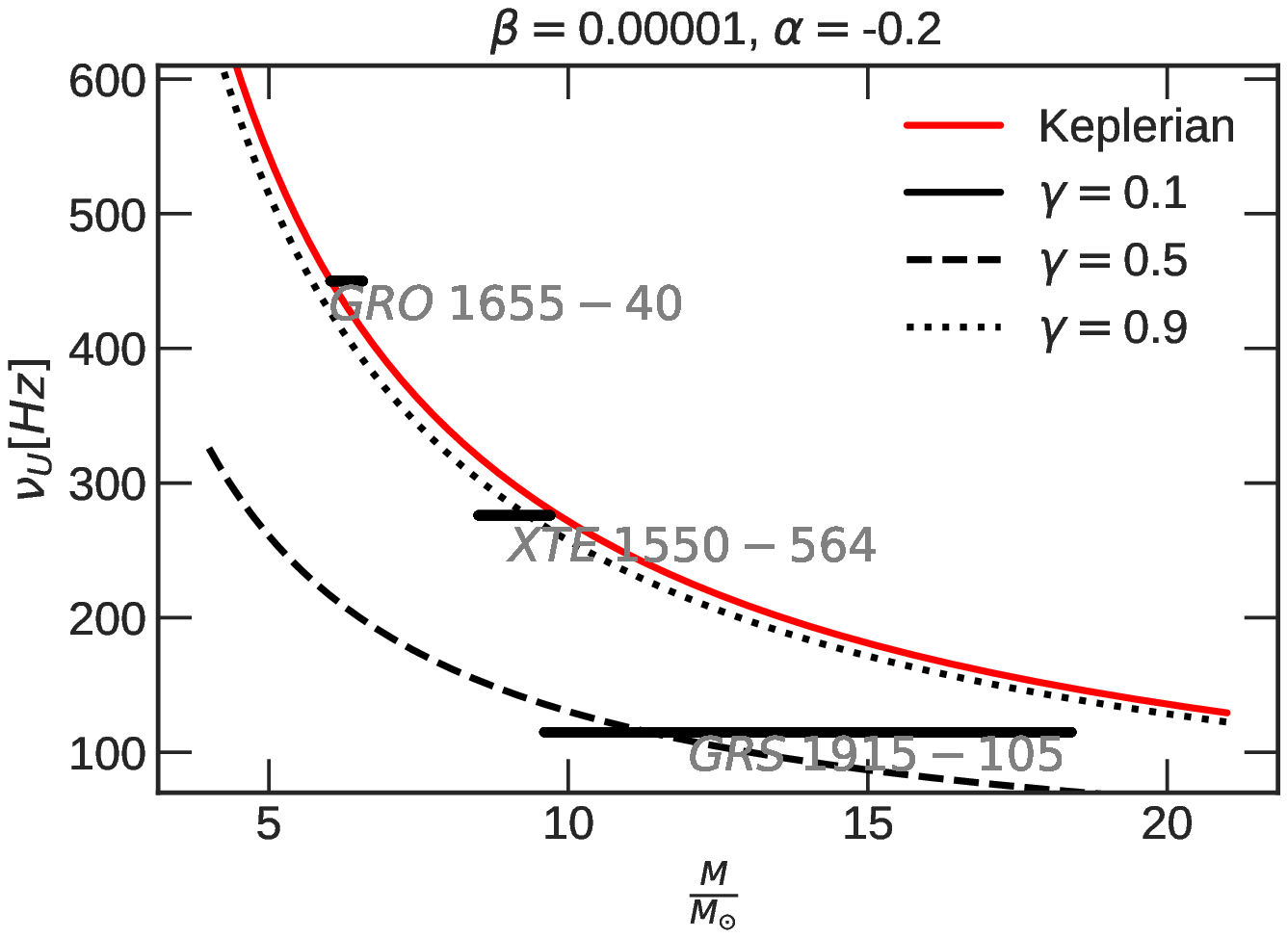} \\
     \includegraphics[width=0.33\hsize]{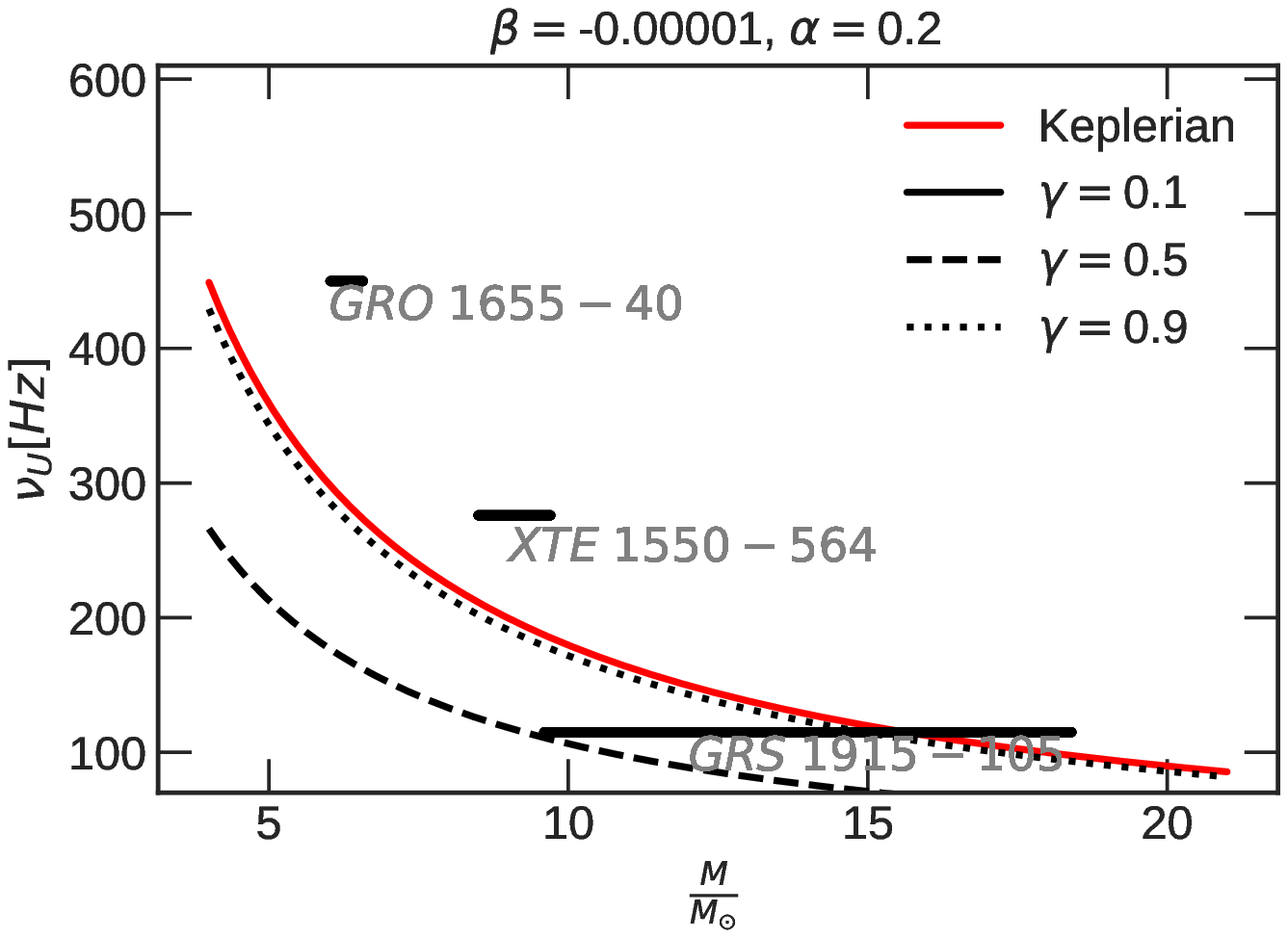} &
     \includegraphics[width=0.33\hsize]{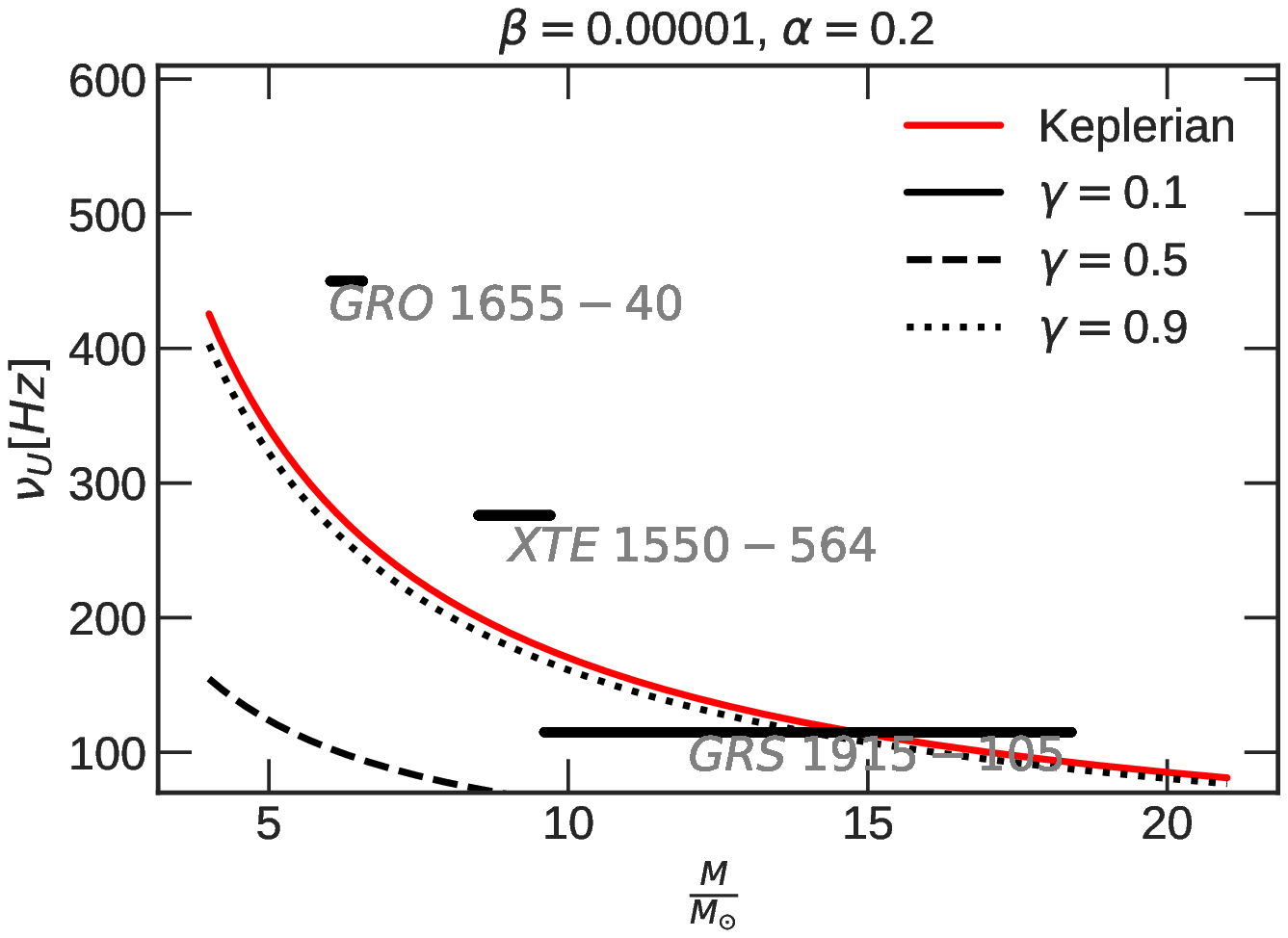} 
     \end{tabular}  
    \caption{{The upper oscillation frequency $\nu_U$ at the resonance radius $3:2$ is presented for various combinations of the studied parameters for the TD and WD models. For the test-particle is depicted in red. The black lines present the fluid for various angular momentum. If the linestyle does not appear, it means that the chosen ratio $3:2$ is not possible.}}  
    \label{fig:TDModel}
\end{figure*}


In this section, we study the epicyclic modes of relativistic tori oscillations in the context of the observed QPO frequencies. One can assume that the two oscillatory modes identified with the observed $3:2$ QPOs are excited at the same radius and physical conditions, which is also valid for a broader class of models \cite{2020A&A...643A..31K}. In this sense, our study is relevant to the consideration of the epicyclic oscillation modes in this context. In principle, different classes of QPO models, assume a relation between the epicyclic frequencies of a particle motion and QPO frequencies  \citep{2005A&A...436....1T,2013A&A...552A..10S,2017A&A...607A..69K,2017EPJC...77..860K}. To summarise them, the eigenfrequencies of two modes in resonance should be equal to the radial epicyclic frequency and the vertical epicyclic or to the Keplerian frequency \citep{2004ragt.meet....1A,2005A&A...436....1T,2011A&A...531A..59T}. The parametric resonance models identify the two observed frequencies of $(\nu_{U}, \nu_{L})$ with the eigenfrequencies of the resonance. However, none of these models is well-matched, especially with the full QPOs amplitudes and the visibility on the source spectral data, in the LMXBs, and this area of research is a non complete task.

In the following, we consider kinematic models RP and TD, and the resonant models WD and Kp. The Relativistic Precession Model (RP) is one of the first attempts proposed in \citep{PhysRevLett.82.17,2002nmgm.meet..426S}. In RP model $\nu_U = \Omega$ and $\nu_{p}:= \nu_L = \Omega - \nu_x$. Their correlations are obtained by varying the radius of the associated circular orbit. It is usually assumed that the variable component of the observed X-ray arises from the motion of “hot-spots” or biting inside the accretion disk on a slightly eccentric orbit. The Epicyclic resonance Model (kp) \citep{2005A&A...436....1T} defines the upper frequency as $\nu_U = \Omega$ and the lower frequency as $\nu_L = \omega_x$. The Tidal Disruption Model (TD), a kinematic model \citep{2008A&A...487..527C,2009A&A...496..307K} follows very similar approach as the RPm. In this model, the QPOs are assumed as a result of tidal disruption of large accreting inhomogeneities. In TDM the frequencies are identified with  $\nu_U = \Omega+\omega_x$ and $\nu_L = \Omega$. The Warped disc Model (WDm) is related to non-axisymmetric modes in a warped accretion disc \citep{2004PASJ...56..905K}. In WDM the frequencies are defined as $\nu_U = 2\Omega-\omega_x$ and  $\nu_L = 2\Omega-2\omega_x$. In more realistic versions of this model, the higher harmonic oscillations are also considered up to the third order, then frequencies like $3\Omega-\omega_x$ possible to consider. The weakness of this model is it considers a somehow exotic disc geometry that causes a doubling of the observed lower QPO frequency.


In the Figure \ref{fig:QianVSKep210}, one of the primary focuses is on the influence of the variation of parameters in the angular momentum distribution on the radial epicyclic frequency of the fluid $\kappa$ and $\Omega$. We can note that when increasing $\gamma$, $\kappa$ for the fluid approach goes closer to the test-particle case. This feature is expected since when $\gamma$ is increasing it approaches the Keplerian Angular momentum. About the $\Omega$ the behaviour is similar. We can add that at some radius (bigger radius when increasing $\beta$) the black lines are decreasing faster than the red ones and in consequence crossing each other.

The second point here is about the ratio between these two frequencies and particularly about the $3:2$ ratio between them. For each model, the chosen upper frequency $\nu_U$ is a function of the inverse of the mass $M$ of the central object scaled by a constant related to $\Omega(r_{\star})$ where $r_{\star}$ is the radius of the ratio between the upper and lower frequency. In the particular case of the $KP$ model, $\nu_U \sim \Omega$ and $\nu_L \sim \kappa$. The vertical line in the figure \ref{fig:QianVSKep210} are corresponding to the radius, $r_{\star}$, of the $3:2$ between $\Omega:\kappa$. For certain values of $\gamma$, depending on the metric parameters, the corresponding vertical lines do not appear on the plots. There are two reasons: (i) the $3:2$ ratio is not possible which can happen for positive values of $\beta$ since for a larger radius $\kappa$ can become negative, and (ii) the radius is far away than the horizontal range. We can see that the vertical line has the same behaviour as the frequencies, the ones related to the fluid are approaching closer and closer to the test particle ones. Besides, $\Omega(r_{\star})$ is bigger for the test particle than the tori. The related $\nu_U$ has the same steepness but shifted. This is shown in the Figure \ref{fig:KPModel} for the KP model and in the Figures \ref{fig:RPmodel} and \ref{fig:TDModel} for the RP and TD model, respectively. In this series of plots, we fit the upper frequency to the data of the three mentioned Microquasars \ref{Quasars}. All along these three figures, the legend follows the same pattern as Figure \ref{fig:QianVSKep210}. We can see that in the Kp model, the combinations of the metric and angular momentum distribution parameters do fitting properly. The results are different when we move to the RP and TD models. For some sets of the various parameters, we have a much better crossing with the three data sets. The effect of the pressure, the angular momentum distribution and the metric parameters can not be decorrelated from each other. From the three Figures \ref{fig:KPModel}-\ref{fig:TDModel} a particular parameter $\alpha$ and $\gamma$ coming from the metric and angular momentum distribution respectively, play a more major role in the fitting.


 \begin{table}

\caption{\label{Quasars}Observed HF QPO data for the three micro-quasars, independent of the HF QPO measurement, and based on the spectral continuum fitting.}
\centering
 \begin{tabular}{c c c c} 
 \hline
 Source &  GRO $1655-40$ & XTE $1550-564$ &  GRS $1915+105$ \\ 
 \hline\hline
  $\nu_U$ & $447 - 453$ & $273 - 279$ &  $165 - 171$ \\ 
 $\nu_L$ & $295 - 305$ & $179 - 189$ & $108 - 118$ \\
 $\frac{M}{{M}_{\odot}}$ & $6.03 - 6.57$ & $8.5 - 9.7$ & $9.6 - 18.4$ \\
 $a$ & $0.65 - 0.75$ &  $0.29 - 0.52$ &  $0.98 - 1$ \\[1ex]
 \hline
 \end{tabular}
\end{table}

\section{Summary and conclusion} \label{summary}

This paper studied the oscillation properties and examined different QPO models of relativistic, non-self-gravitating tori around a distorted deformed compact object up to the quadrupole. This space-time is a generalization of the well-known $\rm q$-metric which is static and axisymmetric and contains two distortion parameters $\beta$ and deformation parameter $\alpha$ which is briefly explained in Section \ref{spacetime}. These two parameters alter the epicyclic frequencies' properties in this background.

The main goal of this paper was to study the radial epicyclic frequency in a perfect fluid disk via a local analysis analytically, and we derived the expression of the radial epicyclic frequency for the tori in the equation \eqref{kappadisk} since it was previously derived for a test particle \cite{2022MNRAS.513.3399F}. To achieve this goal, we employ the vertically integrated technique for tori introduced in \cite{10.1046/j.1365-8711.2003.07023.x}. First, we built the tori with constant and non-constant angular momentum distribution \cite{2009A&A...498..471Q} which is different from what is considered in \cite{10.1046/j.1365-8711.2003.07023.x,2004MNRAS.354.1040M}. We discuss how the oscillation properties with parameters in the distributions of angular momentum.

Furthermore, we examined the possibility of relating oscillatory radial frequencies to the frequencies of the high-frequency quasi-periodic oscillations observed in three microquasars GRS $1915+105$, XTE $1550-564$, GRO $1655-40$, and at their frequency ratio $3:2$. In fact, this set-up may open up a variety of exciting applications in astrophysics due to presenting of quadrupole parameters in the model.

In fact, it is possible from observational data or other analytic set-ups to assign some restrictions on the parameters in this metric. Moreover, the next step of this work would be to explore more about the observational data by considering different inputs like rotation in this model that helps to model a more realistic complex system of astronomical objects. The perturbation study via a global approach and investigation of g-mode and p-mode oscillations in this background could be used to explain the harmonic relations in the observed HF QPOs. Also, adding the strong magnetic field, which influences the metric itself, is the subject of the following work.

\section*{Acknowledgments}
S.F. acknowledge the research training group GRK 1620 "Models of Gravity" and the excellence cluster QuantumFrontiers funded by the Deutsche Forschungsgemeinschaft (DFG, German Research Foundation) under Germany’s Excellence Strategy – EXC-2123 QuantumFrontiers – 390837967. A.T. is supported by the research training group GRK 1620 "Models of Gravity" funded by the DFG.

\bibliographystyle{unsrt}
\bibliography{kappapq}

\end{document}